\documentclass[aps,prb,twocolumn,reprint,superscriptaddress,longbibliography]{revtex4-2}

\usepackage{hyperref}
\usepackage{graphicx}  
\usepackage{bm}        
\usepackage{amssymb}   
\usepackage{xspace}
\usepackage{verbatim}
\usepackage{color}
\usepackage{soul}
\usepackage{subfigure}
\usepackage[export]{adjustbox}
\usepackage{dcolumn}
\usepackage{amsthm,stmaryrd,mathtools}
\usepackage{txfonts}
\usepackage{braket}
\usepackage{amsmath}
\usepackage{xcolor}
\usepackage{array}
\usepackage{multirow}
\usepackage{tabularx}
\usepackage{booktabs}
\usepackage{lipsum}

\begin{document}

\author{Rozhin Yousefjani}%
\email{ryousefjani@uestc.edu.cn}
\affiliation{Institute of Fundamental and Frontier Sciences, University of Electronic Science and Technology of China, Chengdu 610051, China}
\affiliation{Qatar Center for Quantum Computing, College of Science and Engineering, Hamad Bin Khalifa University, Doha, Qatar}

\author{Krzysztof Sacha}%
\affiliation{Instytut Fizyki Teoretycznej, Wydział Fizyki, Astronomii i Informatyki Stosowanej,
Uniwersytet Jagiello\'{n}ski, ulica Profesora Stanisława Łojasiewicza 11, PL-30-348 Krak\'{o}w, Poland, Centrum Marka Kaca, Uniwersytet Jagiello\'{n}ski,
ulica Profesora Stanisława Łojasiewicza 11, PL-30-348 Krak\'{o}w, Poland}

\author{Abolfazl Bayat}%
\affiliation{Institute of Fundamental and Frontier Sciences, University of Electronic Science and Technology of China, Chengdu 610051, China}
\affiliation{Key Laboratory of Quantum Physics and Photonic Quantum Information, Ministry of Education, University of Electronic Science
and Technology of China, Chengdu 611731, China}

\title{Discrete Time Crystal Phase as a Resource for Quantum Enhanced Sensing}

\begin{abstract}
{Discrete time crystals are a special phase of matter in which time translational symmetry is broken through a periodic driving pulse.  Here, we first propose and characterize an effective mechanism to generate a stable discrete time crystal phase in a disorder-free many-body system with indefinite persistent oscillations even in finite-size systems. Then we explore the sensing capability of this system to measure the spin exchange coupling. The results show strong quantum-enhanced sensitivity throughout the time crystal phase. As the spin exchange coupling varies, the system goes through a sharp phase transition and enters a non-time crystal phase in which the  performance of the probe considerably decreases. 
We characterize this phase transition as a second-order type and determine its critical properties through a comprehensive finite-size scaling analysis. 
The performance is independent of the initial states and may even benefit from imperfections in the driving pulse.
A simple set of projective measurements can capture the quantum-enhanced sensitivity.}   
\end{abstract}

\maketitle

\section{Introduction} Symmetry breaking is a fundamental process that shapes our universe, from its early evolution and the formation of elementary particles to various forms of phase transitions in our daily lives. Breaking continuous spatial translation symmetry into a discrete one results in ordinary crystals, where atoms sit in a regular order.
In a seminal work by Wilczek~\cite{wilczek2012quantum}, the idea of breaking continuous time translation symmetry and the formation of time crystals was proposed. While this proposal proved impossible in equilibrium states of time-independent systems with two-body interactions \cite{Bruno2013b,watanabe2015absence,Kozin2019}, the spontaneous emergence of a new periodic motion turned out to be possible in periodically driven systems \cite{Sacha2015,khemani2016phase,else2016floquet}. Breaking discrete time translational symmetry (DTTS) in such systems and forming so-called discrete time crystals (DTC) has become the subject of intensive theoretical \cite{Sacha2015,khemani2016phase,else2016floquet,
yao2017discrete,russomanno2017floquet,ho2017critical,huang2018clean,Matus2019,kshetrimayum2020stark,Estarellas,Maskara2021Discrete,Wang2021,
pizzi2021higher,Collura2022Discrete,Huang2022Discrete,
Bull2022Tuning,Deng2023Using,liu2023discrete,Huang2023Analytical,
Giergiel2023} and experimental \cite{zhang2017observation,
choi2017observation,pal2018temporal,rovny2018observation,
smits2018observation,Randall2021,Kessler2021,xu2021realizing,Kyprianidis2021,Taheri2022,mi2022time,frey2022realization,
Bao2024,Kazuya2024,Liu2024,Liu2024a} 
research (for reviews see \cite{sacha2017time,else2020discrete,khemani2019brief,SachaTC2020,Hannaford2022,Zaletel2023}). In periodically driven systems with a period $T$, DTCs do not correspond to equilibrium states but reveal temporal order where: (i) physical observables evolve with period $gT$ with integer $g{>}1$; (ii) the dynamics are robust against small imperfections in the driving pulse; and (iii) the oscillating behavior persists indefinitely in the thermodynamic limit. The existence of the DTC relies on mechanisms that prohibit the system from absorbing energy from the driving pulse, such as self-trapping, the presence of disorder, gradient magnetic fields, all-to-all or long-range interactions, domain-wall confinement, and quantum scars \cite{sacha2017time,else2020discrete,khemani2019brief,SachaTC2020,Zaletel2023,liu2023discrete,kshetrimayum2020stark,russomanno2017floquet,pizzi2021higher,Collura2022Discrete,Maskara2021Discrete,Deng2023Using,Bull2022Tuning,Huang2022Discrete,Huang2023Analytical}. The study has been also extended to non-Hermitian physics~\cite{yousefjani2024NHDTC}. While major proposals focus on the formation and detection of DTCs, the potential application of this phase of matter is yet to be explored. So far, time crystals have been used for simulating complex systems~\cite{Estarellas}, topologically protected quantum computation~\cite{Bomantara2018Simulation}, designing quantum engines~\cite{Carollo2020Nonequilibrium}, metrology in fully-connected graphs~\cite{lyu2020eternal}, measuring AC fields~\cite{Iemini2023}, and system-environment coupling~\cite{montenegro2023quantum,Cabot2024Continuous}.

Strongly correlated many-body systems have been identified as excellent quantum sensors. 
In particular, various forms of quantum criticality have been used for achieving quantum-enhanced sensitivity beyond the capacity of classical sensors. 
This includes first-order~\cite{raghunandan2018high,heugel2019quantum,yang2019engineering}, second-order~\cite{zanardi2006ground,zanardi2007mixed,gu2008fidelity,zanardi2008quantum,invernizzi2008optimal,gu2010fidelity,gammelmark2011phase,skotiniotis2015quantum,rams2018limits,wei2019fidelity,chu2021dynamic,liu2021experimental,montenegro2021global,mirkhalaf2021criticality,di2021critical,Salvia2023Critical}, dissipative~\cite{fernandez2017quantum,baumann2010dicke,baden2014realization,klinder2015dynamical,rodriguez2017probing,fitzpatrick2017observation,fink2017observation,ilias2022criticality,Ilias2023Criticality,Alipour2014Quantum}, topological~\cite{budich2020non,sarkar2022free,koch2022quantum,yu2022experimental}, Floquet~\cite{mishra2021driving,mishra2022integrable}, and Stark~\cite{he2023stark,yousefjani2023Long,yousefjani2024nonlinearityenhanced} phase transitions. For a reference see~\cite{montenegro2024rev}.
However, the benefits of using criticality-based probes are limited by three major factors: (i) the region over which quantum-enhanced precision is achievable is very narrow; (ii) state preparation, e.g. ground state, near the critical point may require a complex time-consuming procedure; and (iii) the presence of imperfection deteriorates the performance of the sensor. Therefore, any sensing protocol that operates optimally over a reasonably wide region without requiring complex state preparation and being stable against unwanted imperfections is highly desired.

Here, by exploiting the state-of-the-art numerical simulations, we put forward a mechanism for establishing a stable DTC with period-doubling oscillations that persist indefinitely even in finite size systems. 
While the DTC shows strong robustness to a certain value of imperfection in the pulse, it goes through a sharp second-order phase transition as the spin exchange coupling varies. Relying on this transition, we devise a DTC quantum sensor that benefits from multiple features. First, the probe shows extreme sensitivity to the exchange coupling across the whole DTC phase, resulting quantum-enhanced sensitivity. 
Second, the probe performance is independent of the initial state. 
Third, the precision enhances by increasing imperfection in the pulse to a certain value. 
Forth, a simple set of projective measurements allows to capture the quantum-enhanced sensitivity. 
In addition, we also characterize the non-DTC phase observing features of ergodic phase in the thermodynamic limit.

\section{Quantum parameter estimation}
We begin by recapitulating the theory of quantum parameter estimation that aims to infer an unknown parameter $\omega$ in a Hamiltonian of a probe by observing the evolution of the probe's state  $\rho(\omega)$.
The uncertainty in estimating $\omega$, quantified through the standard deviation  $\delta\omega$, is lower bounded by Cram\'{e}r-Rao inequality 
$\delta\omega{\geq}1/\sqrt{\mathcal{F}_{C}(\omega)}$ wherein $\mathcal{F}_{C}(\omega){=}\sum_{r}p_r(\omega) [\partial_\omega\ln p_r(\omega)]^2$ is the Classical Fisher information (CFI). Here $p_r(\omega){=}{\rm Tr} [\rho(\omega)\Pi_r]$ is the outcome probability of measuring the probe using operators $\{\Pi_r\}$. Optimizing the measurement operators leads to 
$\delta\omega{\geq}1/\sqrt{\mathcal{F}_{Q}(\omega)}$, known as the quantum Cram\'{e}r-Rao inequality wherein $\mathcal{F}_{Q}(\omega)$ is the quantum Fisher information (QFI).
For pure states $\rho(\omega){=}|\psi(\omega)\rangle\langle \psi(\omega)|$ the QFI is given by 
$\mathcal{F}_{Q}(\omega){=}4 \big(\langle \partial_{\omega}\psi(\omega) | \partial_{\omega}\psi(\omega) \rangle{-}|\langle \partial_{\omega}\psi(\omega) | \psi(\omega) \rangle|^2 \big)$ \cite{fisher1922mathematical}.
In classical sensors Fisher information, at best, scales linearly with system size $L$. Exploiting quantum features in sensing the coupling of a $k$-body interacting system allows precision enhancement to $\mathcal{F}_Q{\sim} L^{2k}$, known as ultimate precision~\cite{boixo2007generalized}.

\section{The model} 
We consider a one-dimensional chain that contains $L$ spin-$1/2$ with Ising-type interaction, governed by the following Hamiltonian
\begin{align}\label{Eq.Hamiltonian}
H(t) = J H_{I} + \sum_n \delta(t-nT)H_{P}, 
\cr
H_{I} = \sum_{j=1}^{L-1}j\sigma^{z}_{j}\sigma^{z}_{j+1}, \quad  H_{P}  = \Phi \sum_{j=1}^{L}\sigma^{x}_{j}.
\end{align}
Here $J$ is the spin exchange coupling, and $\sigma^{x,y,z}_{j}$ are the Pauli operators. 
The gradient $zz$ interaction in $H_I$ causes off-resonant energy splitting at each site and, therefore, leads the particle's wave function to localize, reminiscent of the  localization which is usually induced by applying a gradient magnetic field~\cite{schulz2019stark,morong2021observation,he2023stark,yousefjani2023Long,yousefjani2024nonlinearityenhanced}. 
This localization, characterized by the existence of an extensive number of conserved quantities~\cite{alet2018many,luitz2015many,Rozha1}, is essential to prevent our system from absorbing the energy of the periodic drives~\cite{lenarvcivc2020critical,Rozha2}.
In the absence of the localization, any local physical observable becomes featureless, and the system thermalizes~\cite{d2014long,lazarides2014equilibrium}.
Since $H_P$ acts in period $T$, the Floquet unitary operator for one-period evolution is
\begin{equation}\label{Eq.FloquetUnitary}
U_{F}(\Omega,\varepsilon) = e^{-iH_{P}} e^{-i\Omega H_{I}},    
\end{equation}
here $\Omega{=}JT$, and $\Phi$  is tuned to be $\Phi{=}(1{-}\varepsilon)\frac{\pi}{2}$, with $\varepsilon$ as deviation from a $\pi/2$ $x$-rotation.
In the following, we show how two main parameters, namely $\Omega$ and $\varepsilon$ play roles in establishing a stable DTC. 
First, we analytically show that setting $\Omega{=}\pi/2$ results in a stable period doubling  DTC that is robust against arbitrary imperfection $\varepsilon$.
Then, through comprehensive numerical simulations, we show that as $\Omega$ varies from $\pi/2$, the system goes through a sharp phase transition from a stable  DTC  to a regime in which DTC order is lost.
We explore the possibility of this phase transition to act as a resource for quantum sensing.

\begin{figure}
    \centering
    \includegraphics[width=0.49\linewidth]{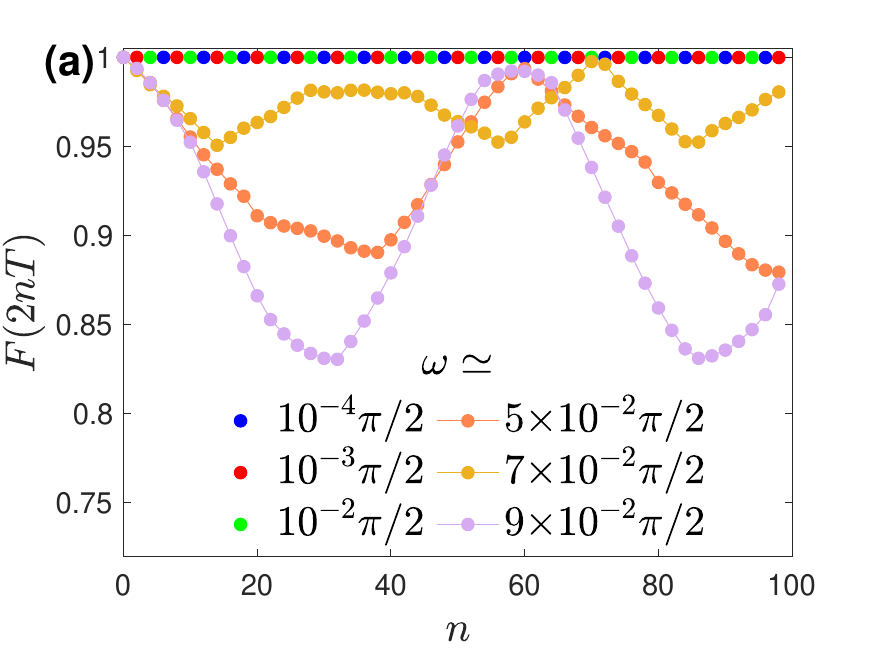}
    \includegraphics[width=0.49\linewidth]{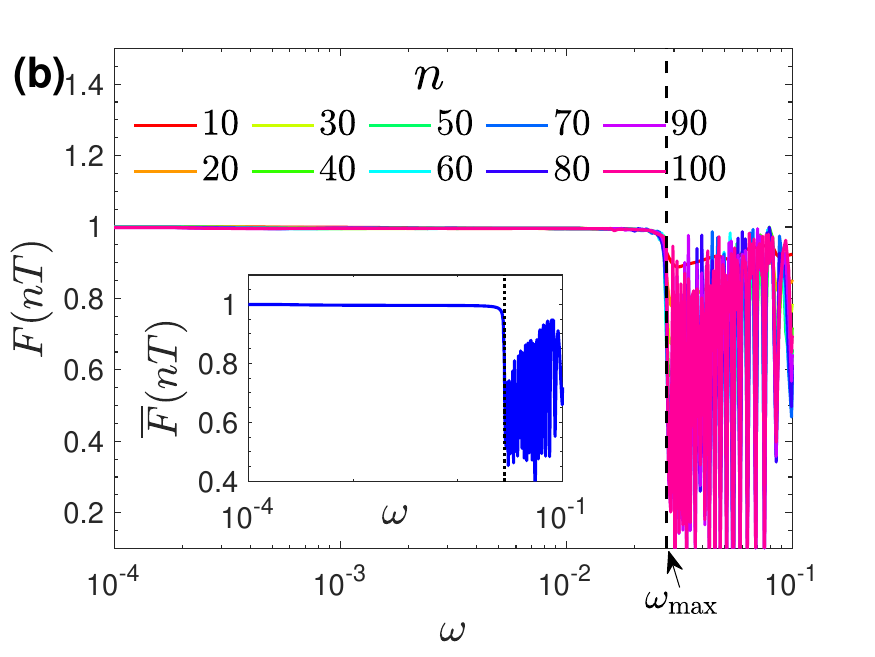}
    \caption{(a) Stroboscopic dynamics of the revival fidelity $F(2nT)$ over a hundred of period cycles $n$ when system of size $L{=}12$ is in the  DTC phase, happens for $\omega{\leq}10^{-2}\pi/2$, and non-DTC, happens for $\omega{\geq}0.05\pi/2$. (b) Dynamical behavior of the revival fidelity at stroboscopic times $n{\in}\{10,\cdots,100\}$ as a function of $\omega$ obtained for a chain of size $L{=}30$ and $\varepsilon{=}0.01$. The inset is the average fidelity $\overline{F}(nT)$. The black dashed line determines the onset of the phase transition. }
    \label{fig:Fig1}
\end{figure}

\section{ Discrete Time Crystal} 
We begin by highlighting some key features of $H_{I}$.
First, $H_I$ is diagonalized in the computational basis $\{|\mathbf{z}\rangle\}$, namely $H_{I}{=}\sum_{\mathbf{z}=1}^{2^L}E_\mathbf{z} |\mathbf{z}\rangle \langle \mathbf{z}|$.
Here
$|\mathbf{z}\rangle{=}(\sigma_{1}^{x})^{j_1}(\sigma_{2}^{x})^{j_2}\cdots(\sigma_{L}^{x})^{j_L}|{\uparrow,\uparrow,\cdots,\uparrow}\rangle$ with  $\mathbf{z}{=}(j_1,j_2,\cdots,j_L)_2$ being the binary representation of the integer $\mathbf{z}$.
Second, $[H_{I},\Pi_{j}\sigma_{j}^{x}]{=}0$
which implies that $E_\mathbf{z}{=}E_{2^L-1-\mathbf{z}}$. 
Third, for an even number of spins which is considered here, all the eigenvalues $E_{\mathbf{z}}$ are integer numbers 
that are even (odd) if $L{/2}$ is an even (odd) number.  
Since $[H_{I},\Pi_{j}\sigma_{j}^{x}]{=}0$,  for $\varepsilon{=}0$ which results in $e^{-iH_{P}}{=}\Pi_{j}\sigma_{j}^{x}$, one has a trivial period doubling DTC as
$U_F^{2}(\Omega{=}\pi/2,\varepsilon{=}0){=}e^{-2iH_{P}} e^{-2i\Omega H_{I}}{=}(-1)^{H_{I}}{=}{\pm}\mathbb{I}$. 
Consequently, one observes persistent oscillations in typical observables with spontaneously breaking DTTS.
For $\varepsilon{\neq}0$, one gets $[H_{I},e^{-iH_{P}}]{\neq}0$. In this case, the reduction of $U_{F}^2(\Omega{=}\pi/2,\varepsilon{\neq}0)$ to the identity is not obvious. 
To study this nontrivial DTC, we focus on dynamic of $|\mathbf{z}\rangle$ over $n$ period cycles and its revival fidelity $F(nT){=}|\langle \mathbf{z}|U_F^{n}(\Omega{=}\pi/2,\varepsilon{\neq}0)|\mathbf{z}\rangle|^2$.
For a typical $|\mathbf{z}\rangle$ with $u_{\mathbf{z}}$ spins down, one has $e^{-i\Omega H_{I}}|\mathbf{z}\rangle{=}(-i)^{L/2}(-1)^{u_{\mathbf{z}}} |\mathbf{z}\rangle$.
Then the first rotating pulse evolves $|\mathbf{z}\rangle$ to a superposition of all the $2^{L}$ elements, each with coefficient $(-i\sin\Phi)^{f}(\cos\Phi)^{L-f}$ wherein $f$ is the number of the flipped spins.
After the second period of the evolution,  
one can show that $\langle \mathbf{z}|U_F^{2}(\Omega{=}\pi/2,\varepsilon{\neq}0)|\mathbf{z}\rangle$ is the summation of $2^{L}$ choices of flipping $L$ spins with coefficient $(i)^{L}(-i\sin\Phi)^{2f}(\cos\Phi)^{2(L-f)}$. A straightforward simplification results in  $\langle \mathbf{z}|U_F^{2}(\Omega{=}\pi/2,\varepsilon{\neq}0)|\mathbf{z}\rangle{=}(-i)^{L}$ and, hence, $F(2T){=}1$, see appendix A. 
This evidences that regardless of the imperfections $\varepsilon$, as long as $\Omega{=}\pi/2$ any initial state returns to itself after $2T$, therefore, period-doubling oscillations of $F(nT)$ persists indefinitely even in finite size systems. 
\\

However, establishing a stable DTC must be independent of fine-tuned Hamiltonian parameters.
This obliges us to analyze the effect of a deviation as $\omega{=}|\pi/2{-}\Omega|$.
Surprisingly, our comprehensive numerical simulations show that as $\omega$ increases, our system goes through a sharp phase transition from a stable DTC to a region with no spontaneous breaking of DTTS in Eq.(\ref{Eq.Hamiltonian}). 
Before detailing the main findings, some methodological notes must be clarified.
We used the exact diagonalization (ED) for $L{=}12$ and time-dependent variational principle (TDVP) techniques for finite matrix product state (MPS), using PYTHON package TeNPy~\cite{tenpy}, for $L{>}12$.
The results are presented for the initial state $|\psi_{0}\rangle{=}|\mathbf{0}\rangle{=}|{\uparrow,\cdots,\uparrow\rangle}$, although, the results are generic and remain valid for other computational basis states too (see appendix B).
In Fig.~\ref{fig:Fig1}(a), we plot stroboscopic dynamics of the revival fidelity $F(2nT)$ as a function of $n$ for various $\omega$ in the system of size $L{=}12$ under a driving pulse with $\varepsilon{=}0.01$.
In the stable DTC phase, happening in the range $\omega{\leq}10^{-2}\pi/2$, one observes $F(2nT){=}1$. For larger values of the deviation, such as $\omega{\geq}0.05\pi/2$, revival fidelity shows nontrivial oscillations, signaling the entrance to a non-DTC region. We characterize this region later.
This distinctive behavior with respect to $\omega$ reflects itself in all stroboscopic times as has been depicted in Fig.~\ref{fig:Fig1}(b). In this panel, we plot the revival fidelity at different stroboscopic times $n{\in}\{2,10,\cdots,100\}$, in a chain of length $L{=}30$ and $\varepsilon{=}0.01$. 
The inset represents the average fidelity $\overline{F}(nT){=}(1/N)\sum_{n=1}^{N}F(2nT)$ for the considered stroboscopic times.
As is obvious from Fig.~\ref{fig:Fig1}(b), the phase transition between stable DTC  and non-DTC region occurs at a specific value of $\omega{=}\omega_{\max}$, dashed line, in all the stroboscopic times.
In the following, we first analyze the capability of this phase transition as a resource for quantum sensing. Then, we complete this analysis by extracting the critical features of the quantum phase transition using a well-established mechanism that identifies the type of transition as a second-order one.

\section{ DTC sensor} 
\begin{figure}
    \centering
    \includegraphics[width=0.49\linewidth]{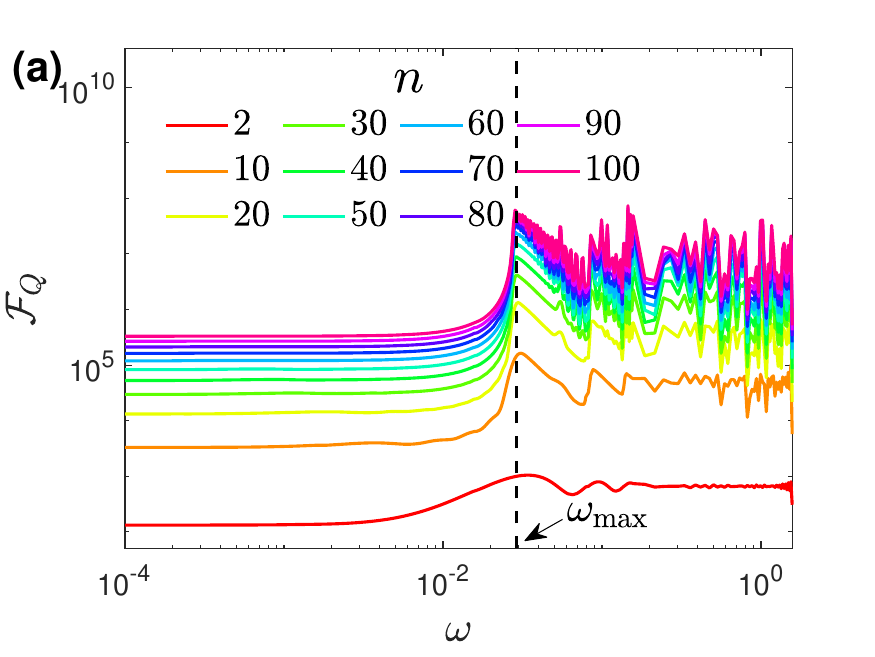}
    \includegraphics[width=0.49\linewidth]{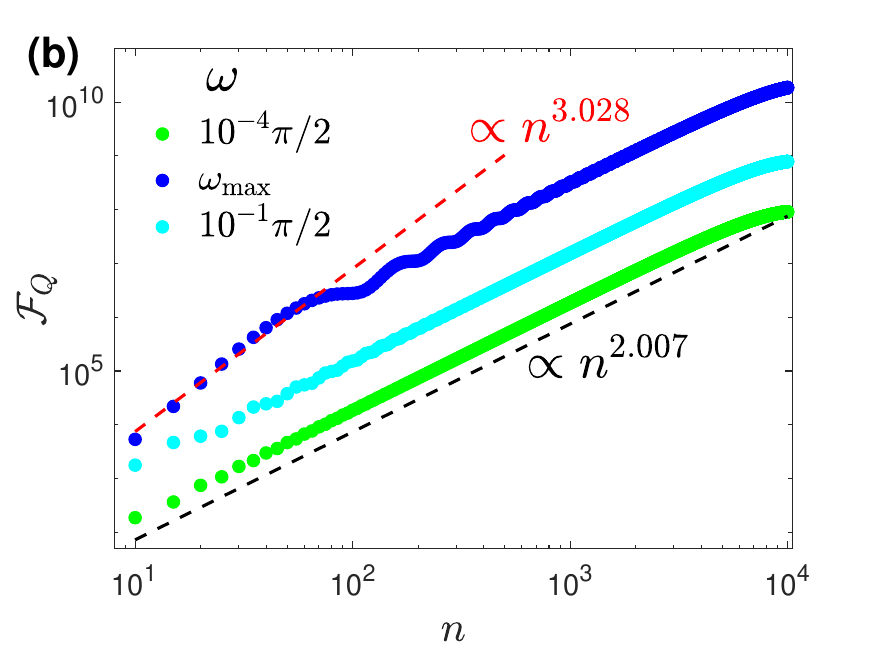}
    \includegraphics[width=0.49\linewidth]{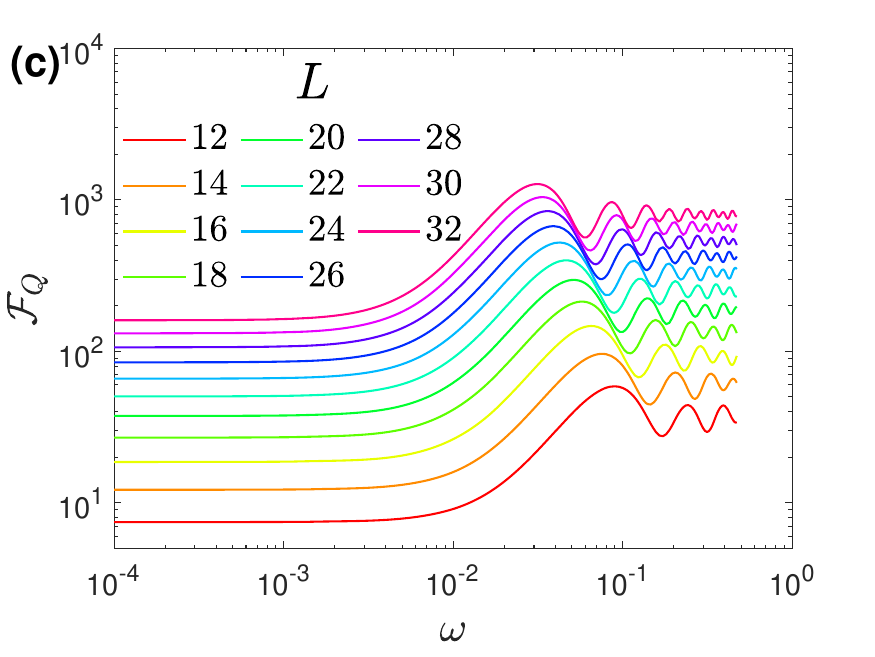}
    \includegraphics[width=0.49\linewidth]{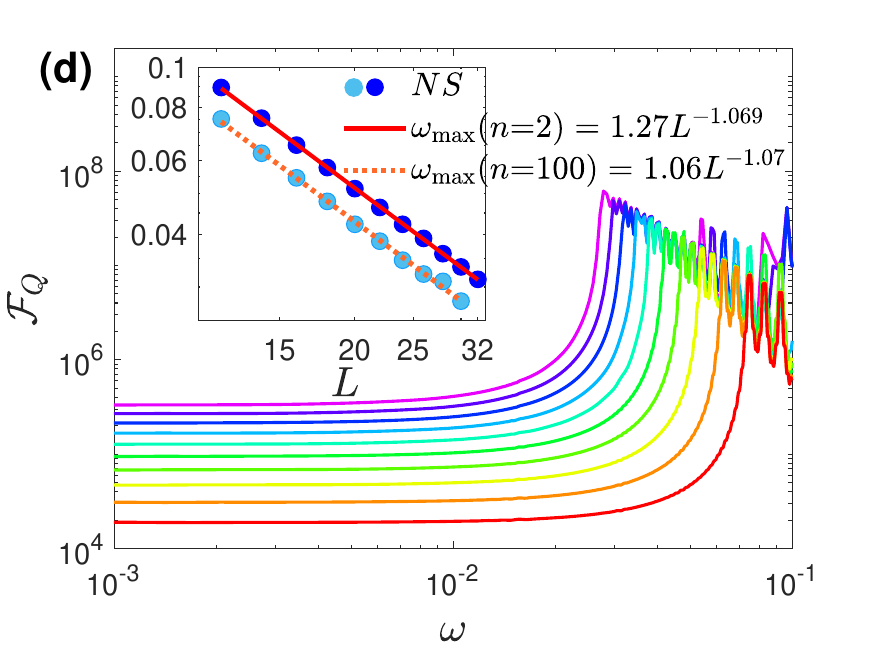}
    \includegraphics[width=0.49\linewidth]{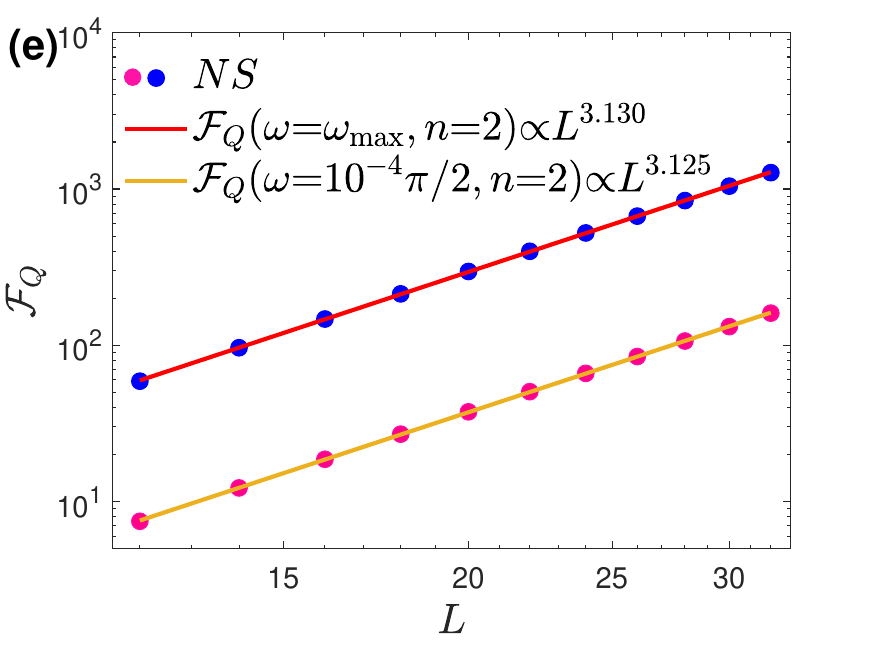}
    \includegraphics[width=0.49\linewidth]{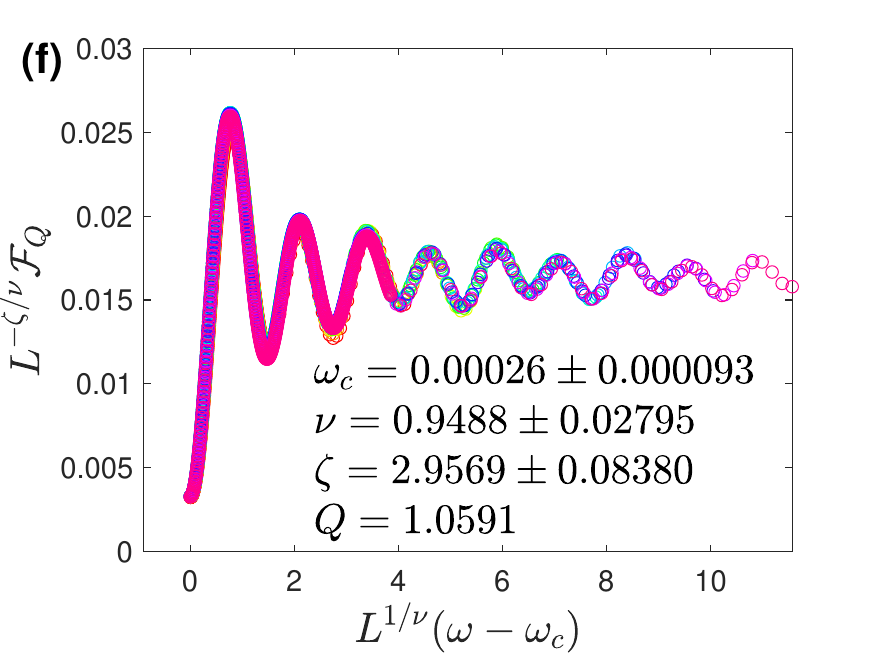}
    \caption{(a) The QFI $\mathcal{F}_{Q}$ versus $\omega$ in stroboscopic times in a system of size $L{=}30$. The onset of the phase transition is determined by  $\omega{=}\omega_{\max}$ (dashed line), the point where QFI peaks in different $n$'s. (b) Dynamical growth of the QFI when a system of size $L{=}12$ is deeply in DTC phase (for $\omega{=}10^{-4}\pi/2$), in the non-DTC phase (for $\omega{=}10^{-1}\pi/2$), and at the transition point ($\omega{=}\omega_{\max}$). (c) and (d) QFI versus $\omega$ in systems with various $L$'s after $n{=}2$ and $n{=}100$, respectively. Inset: $\omega_{\max}$ versus $L$ at $n{=}2$ and $n{=}100$. The numerical simulation (NS) are well described by the fitting function $\omega_{\max}{\propto}L^{-1}$. (e) The values of the QFI after $n=2$ in DTC phase (for $\omega{=}10^{-4}\pi/2$) and at transition points ($\omega{=}\omega_{\max}$) versus $L$. The numerical simulation (NS) is well-mapped by a function as $\mathcal{F}_{Q}{\propto}L^{\beta}$ (solid lines) with $\beta{>}3$. (f) The finite-size scaling analysis obtained for the curves in panel (c). The best data collapse is obtained for the reported critical parameters ($\omega_{c},\zeta,\nu$). Here, $Q$ determines the quality of the data collapse with $Q{=}1$ for the optimal data collapse~\cite{melchert2009autoscalepy,andreas_sorge_2015_35293}. }\label{fig:Fig2}
\end{figure}
To investigate the sensing capability of our  DTC probe for sensing $\omega$, in Fig.~\ref{fig:Fig2}(a), we plot QFI $\mathcal{F}_{Q}$ as a function of $\omega$ at different stroboscopic times $n{\in}\{2,10,\cdots,100\}$, in a chain of length $L{=}30$ and $\varepsilon{=}0.01$.
Several interesting features can be observed. 
First, the QFI shows distinct behaviors in each phase. 
While in the DTC phase, the $\mathcal{F}_{Q}$ becomes a plateau whose value depends on $n$, in the non-DTC region it shows nontrivial and fast oscillations.
Second, by approaching the transition point, denoted by $\omega_{\max}$ (dashed line), the QFI indeed shows a clear peak at all stroboscopic times.
Note that $\omega_{\max}$ in both Fig.~\ref{fig:Fig1}(b) and Fig.~\ref{fig:Fig2}(a) are exactly the same.
To understand the dynamical growth of the QFI, in Fig.~\ref{fig:Fig2}(b), we plot $\mathcal{F}_{Q}$ over thousands of driving cycle $n$ in a systems of size $L{=}12$ and $\varepsilon{=}0.01$ at different $\omega$'s.
Clearly, when the probe is tuned to work deeply in either DTC phase (for $\omega{=}10^{-4}\pi/2$) or the non-DTC region (for $\omega{=}10^{-1}\pi/2$), one obtains $\mathcal{F}_{Q}{\propto}n^{2}$. However, in the transition point, $\omega{=}\omega_{\max}$ the QFI in the early times $n{\in}[1,\cdots,80]$ dramatically increases as $\mathcal{F}_{Q}{\propto}n^{3}$, and then follows $\mathcal{F}_{Q}{\propto}n^{2}$ in the larger times.
To identify the effect of size on quality of sensing, we analyze the QFI at various sizes $L{=}12,\cdots,32$ and also different cycles $n$.
In Figs.~\ref{fig:Fig2}(c) and (d), we plot $\mathcal{F}_{Q}$ as a function of $\omega$ after $n{=}2$ and $n{=}100$ cycling periods, respectively, for various $L$ and fixed $\varepsilon{=}0.01$.
The finite-size effect is obvious in both DTC phase and transition point. 
By enlarging the chain, the peaks of the QFI smoothly skew towards smaller $\omega$, see the inset of Fig.~\ref{fig:Fig2}(d). 
The obtained $\omega_{\max}$ at different $n$'s are well-mapped with function $\omega_{\max}{\propto}L^{-1}$, indicating that in the thermodynamic limit $L{\rightarrow}\infty$ the transition happens at infinitesimal deviation $\omega$.
In the non-DTC region, QFI oscillations, especially over extended periods ($n{=}100$), hinder scaling behavior investigation.
In Fig.~\ref{fig:Fig2}(e), we present the QFI at $n{=}2$ as a function of $L$  at $\omega{=}10^{-4}\pi/2$, namely deep inside the DTC phase, and also at the corresponding transition points $\omega{=}\omega_{\max}$.
The numerical results can be properly mapped with a fitting function as $\mathcal{F}_{Q}{\propto}L^{\beta}$ with $\beta{=}3.125$ and $\beta{=}3.13$ in the DTC phase and at the transition point, respectively. Based on these,  one can suggest the following ansatz for the QFI 
\begin{equation}
\mathcal{F}_{Q} \propto n^\alpha L^\beta,  
\end{equation}
where throughout the DTC phase one has $\alpha {\simeq} 2$ and $\beta {\simeq} 3$.  
We highlight this as the main result of this Letter showing that our DTC probe achieves quantum-enhanced sensitivity. It is worth emphasizing that in classical probes one at best achieves $\beta{=}1$. 
Exploiting quantum features may enhance the precision to $\beta{=}2$, known as the Heisenberg limit.
The ultimate precision in $k$-body interacting systems is $\beta{=}2k$, which equals $\beta{=}4$ in our case~\cite{boixo2007generalized}.  

To study the observed phase transition, we start with a continuous second-order ansatz for the QFI as $\mathcal{F}_{Q}{=}L^{\zeta/\nu} \mathcal{G}(L^{1/\nu}(\omega{-}\omega_{c}))$ where $\zeta$ and $\nu$ are critical exponents, $\omega_{c}$ is the critical point and $\mathcal{G}$ is an arbitrary function. If this ansatz is correct, one expects to obtain data collapse of various size systems when $L^{-\zeta/\nu}\mathcal{F}_{Q}$ is plotted versus $L^{1/\nu}(\omega{-}\omega_c)$. Indeed, as shown in Fig.~\ref{fig:Fig2}(f), tuning the parameters to $(\omega_{c},\zeta,\nu){\simeq}(0.00026,2.9569,0.9488)$, optimized using Python package PYFSSA~\cite{melchert2009autoscalepy,andreas_sorge_2015_35293}, results in an almost perfect data collapse for curves in Fig.~\ref{fig:Fig2}(c). This indicates that the  DTC phase transition is indeed of the second-order type. 

\section{Melting transition of the  DTC} 
\begin{figure}
    \centering
    \includegraphics[width=0.49\linewidth]{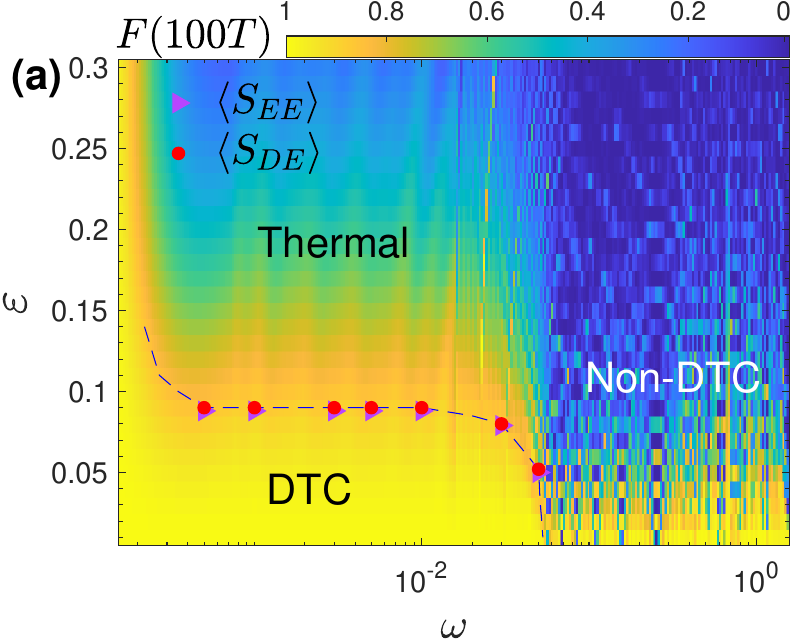}
    \includegraphics[width=0.49\linewidth]{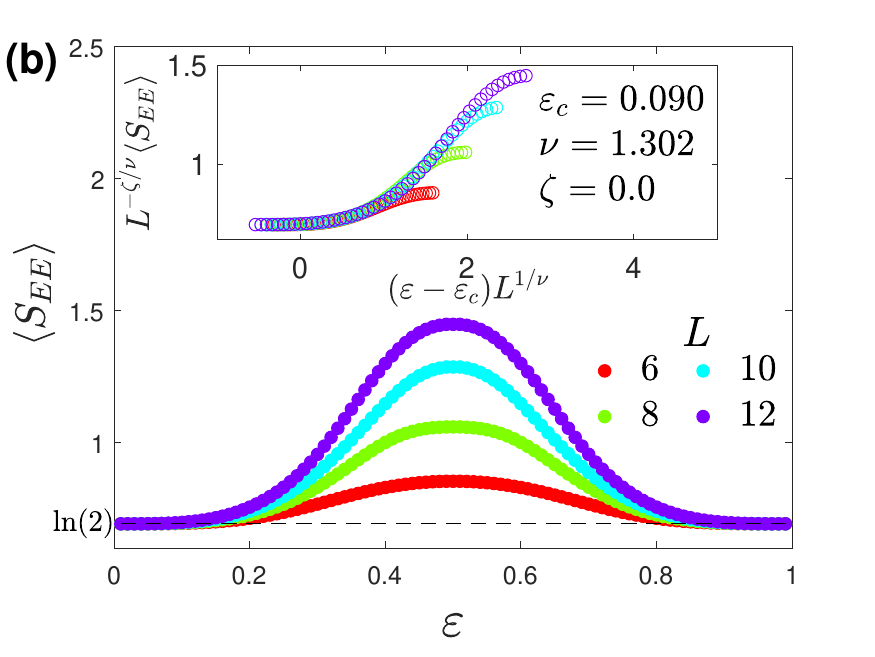}
    \caption{(a) Phase diagram of the  DTC as function of $\omega$ and $\varepsilon$. The background is $F(nT)$ after $n{=}100$ in a system of size $L{=}12$.  The markers determine the phase boundary between  DTC and the thermal phase, obtained through finite-size scaling analysis of averaged entanglement entropy $\langle S_{EE} \rangle$ and averaged diagonal entropy $\langle S_{DE} \rangle$. (b) $\langle S_{EE} \rangle$ versus $\varepsilon$ for systems of different sizes at $\omega{=}5{\times}10^{-4}$. Inset is the finite-size scaling analysis, showing the best data collapse obtained for reported critical parameters~$(\varepsilon_{c},\nu,\zeta)$.    }\label{fig:Fig3}
\end{figure}

Having elucidated the sharp second-order phase transition controlled by $\omega$, we now explore the melting of the  DTC by increasing the imperfection $\varepsilon$. 
In Fig.~\ref{fig:Fig3}(a), we plot the revival fidelity $F(nT)$ after $n{=}100$ as a function of $\omega$ and $\varepsilon$, for system of size $L{=}12$. 
Indeed the phase diagram is fully described by $\omega$ and $\varepsilon$. 
To diagnose the transition driven by the imperfection $\varepsilon$, we use averaged entanglement entropy $\langle S_{EE} \rangle$ and averaged diagonal entropy $\langle S_{DE} \rangle$  obtained for all Floquet states of $U_{F}(\Omega,\varepsilon){=}\sum_{k=1}^{2^L}e^{-i\phi_k}|\phi_{k}\rangle\langle \phi_k|$. For a given Floquet state $|\phi_{k}\rangle$, the reduced density matrix $\rho_{L/2}^{(k)}=\mathrm{Tr}_{L/2} |\phi_{k}\rangle \langle \phi_{k}|$ can be obtained by tracing out $L/2$ spins in the right side of the chain. 
Therefore, the entanglement entropy between the half-systems is $S_{EE}^{(k)}{=}{-}\mathrm{Tr}[\rho_{L/2}^{(k)}\ln(\rho_{L/2}^{(k)})]$ with an average as $\langle S_{EE} \rangle{=}\sum_{k=1}^{2^L}S_{EE}^{(k)}/2^L$.  
Replacing $\rho_{L/2}^{(k)}$ by decohered density matrix $\varrho_{L/2}^{(k)}$ which only contain the diagonal elements of $\rho_{L/2}^{(k)}$, results in diagonal entropy as $S_{DE}^{(k)}{=}{-}\mathrm{Tr}[\varrho_{L/2}^{(k)}\ln(\varrho_{L/2}^{(k)})]$ and its average $\langle S_{DE} \rangle{=}\sum_{k=1}^{2^L}S_{DE}^{(k)}/2^{L}$.
This quantity has recently been proposed for emulating the thermodynamic behavior in many-body localization contexts~\cite{Rozha1,Rozha2}.
In the DTC phase, each Floquet state is a maximally entangled Greenberger-Horne-Zeilinger (GHZ) of two computational basis states.
For instance, for negligible $\varepsilon$, one has $|\phi_{1}\rangle{\cong}\frac{1}{\sqrt{2}}(|\mathbf{0}\rangle {+}|{\mathbf{2^{L}-1}}\rangle)$ and $|\phi_{2^L}\rangle{\cong}\frac{1}{\sqrt{2}}(|\mathbf{0}\rangle {-} |\mathbf{2^{L}-1}\rangle)$ with the corresponding eigenvalues as $\phi_{1}{\cong}E_{\mathbf{1}}$ and $\phi_{2^L}{\cong}E_{\mathbf{1}}\pm \pi$.
This so-called $\pi$-pairs of the Floquet states results in $S_{EE}^{(k)}{=}S_{DE}^{(k)}{=}\ln 2$ in   deep  DTC phase. By increasing $\varepsilon$ both entanglement and diagonal entropy grow and peak at $\varepsilon{=}0.5$, see Fig.~\ref{fig:Fig3}(b) and appendix C.
To characterize critical properties, finite-size scaling analysis needs to be established.
In the inset of Fig.~\ref{fig:Fig3}(b), we depict the best collapse of the corresponding curves obtained for reported $(\varepsilon_{c},\nu,\zeta)$. 
By increasing length $\omega_{\max}$ decreases and thus the extension of the DTC phase becomes smaller. Therefore, for finite-size scaling analysis we select the lengths such that for the given $\omega$ they are all within the DTC phase when $\varepsilon{\simeq}0$.
Note that the obtained $\varepsilon_{c}$ from finite-size scaling of both $\langle S_{EE} \rangle$ and $\langle S_{DE} \rangle$, shown as markers on panel (a), are very close and determine the phase boundary between the DTC and the non-DTC phase.
Note that, while strong imperfection, i.e. large $\varepsilon$, melts the DTC, small imperfections may indeed enhance the sensitivity of our probe, see Appendix C.

\section{Non-DTC region} 
\begin{figure}
    \centering
    \includegraphics[width=0.5\linewidth]{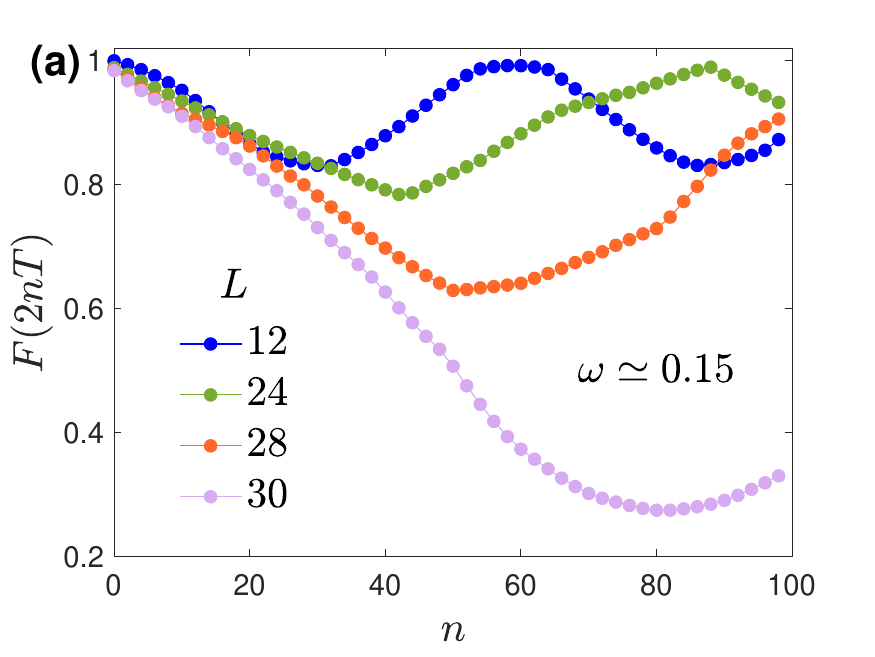}
    \includegraphics[width=0.49\linewidth]{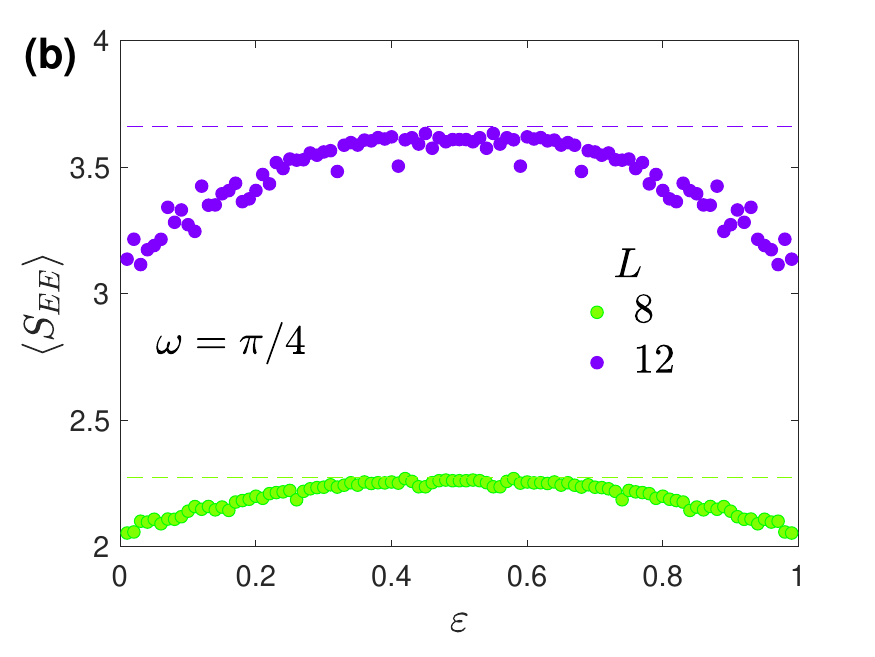}
    \includegraphics[width=0.49\linewidth]{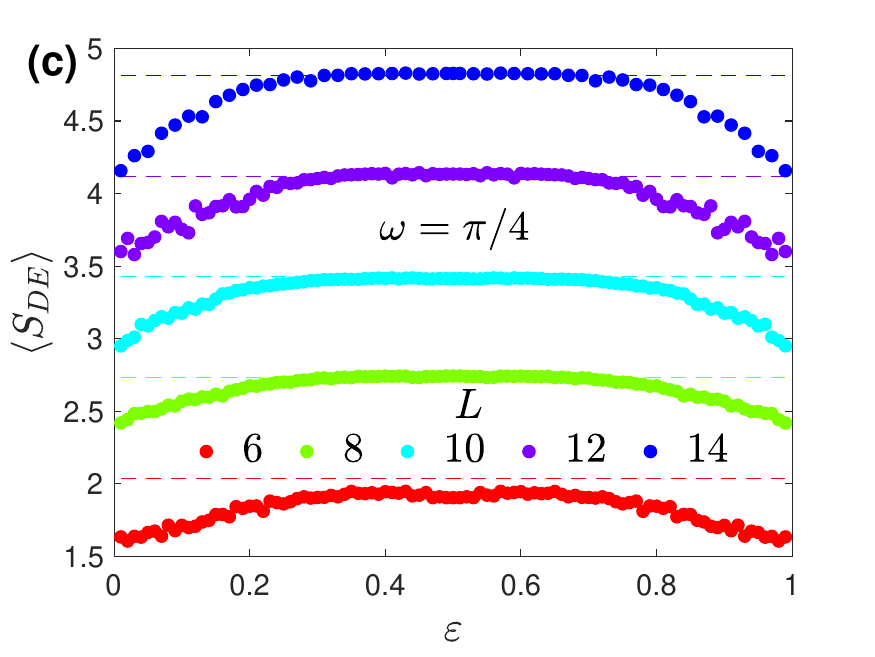} 
    \caption{(a) the stroboscopic dynamics of the revival fidelity $F(2nT)$ over $100$ period cycles when systems with different sizes evolve in non-DTC region with $\omega{\simeq}0.15$ and $\varepsilon{=}0.01$.  (b) the averaged entanglement entropy $\langle S_{EE} \rangle$ and (c) the averaged diagonal entropy $\langle S_{DE} \rangle$ as a function of pulse imperfection $\varepsilon$ obtained for systems of different sizes and $\omega{=}\pi/4$. Dashed lines in (b) and (c) determine the corresponding $\langle S^{P}_{EE} \rangle$ and $\langle S^{P}_{DE} \rangle$, respectively.}
    \label{fig:FigS4}
\end{figure}

By increasing the divination $\omega$, 
the perfect and stable revivals of the fidelity in stroboscopic times, namely $F(2nT){=}1$ for $\omega{\leq} \omega_{\max}$, are replaced by nontrivial oscillations for $\omega{>} \omega_{\max}$. 
This hints one enters a non-DTC region. 
In this section, we aim to characterize the nature of this region.
Our results for the revival fidelity $F(2nT)$ as a function of $n$ for systems of different sizes that are tuned to work in the non-DTC region, namely for $\omega{\simeq}0.15$, have been illustrated in Fig.~\ref{fig:FigS4}(a). 
By enlarging the system size, the period of these incommensurate fluctuations increases. This implies that, in systems with enough large sizes, these oscillations practically vanish in a reasonable time window, signaling the thermalization of the system.
This observation receives more support from our static study based on the entanglement entropy and diagonal entropy. 
In a thermal system, the Floquet states $\{|\phi_{k}\rangle\}$ are expected to behave as a typical random pure state, therefore their entanglement entropy is predicted to follow the Page entropy $\langle S^{P}_{EE} \rangle{\simeq}(L\ln(2)-1)/2$  for enough large $L$'s~\cite{page1993average,torres2017extended,torres2016realistic}. 
In this case, the average entanglement entropy should already captured its maximum and the variations of $\varepsilon$ may not considerably affect $\langle S^{P}_{EE} \rangle$. 
Our numerical results in Fig.~\ref{fig:FigS4}(b) support this prediction for $\omega{=}\frac{\pi}{4}$, namely deep inside the thermal phase. 
The results for systems of size $L{=}8$ and $L{=}12$ capture the Page entropy $\langle S_{EE} \rangle{\simeq}\langle S^{P}_{EE} \rangle$ (depicted by colored dashed lines) with slight changes in terms of $\varepsilon$. 
Regarding the diagonal entropy, typical random pure states are expected to follow $\langle S^{P}_{DE} \rangle{\simeq}\ln(0.48 \times 2^{L/2}) + \ln(2)$.
The presented results in Fig.~\ref{fig:FigS4}(c) confirm this behavior.
In this panel $\langle S^{P}_{DE} \rangle$ is represented by colored dashed lines.

\section{Experimental realization}
\begin{figure}
    \centering
    \includegraphics[width=0.49\linewidth]{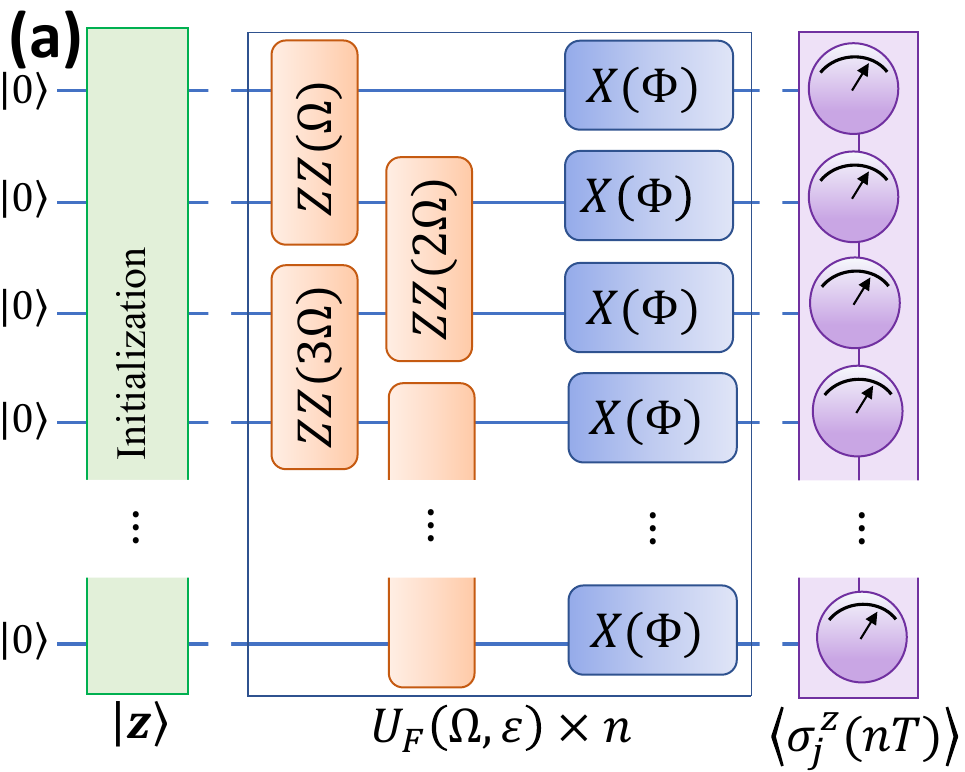}
    \includegraphics[width=0.49\linewidth]{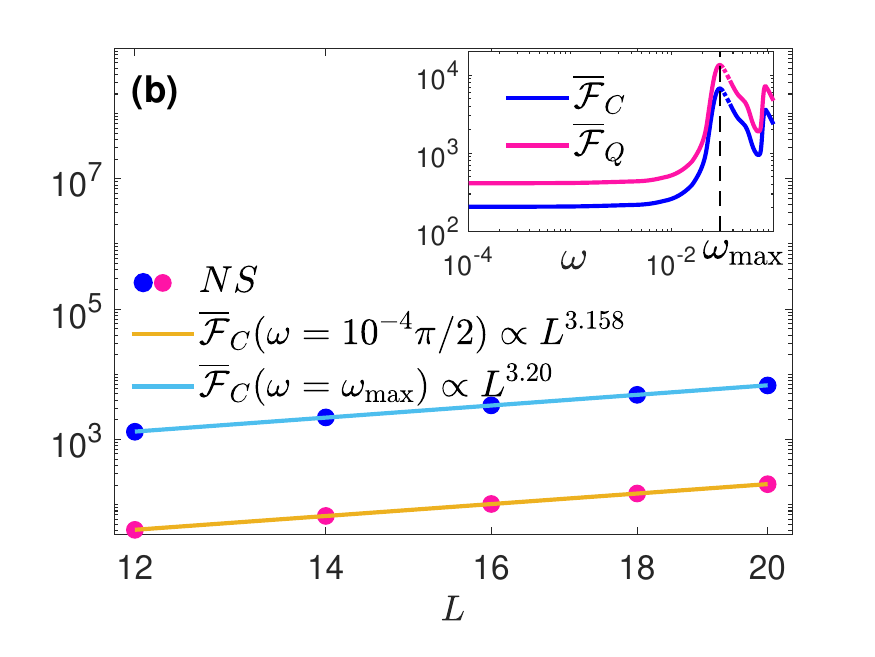}
    \caption{(a) Quantum circuit for implementing our DTC on a digital quantum simulators. The operator $U_F(\Omega,\varepsilon)$ is realized through a combination of CPHASE gates and  $ZZ(j\Omega)$ and single-qubit rotations $X(\Phi)$. (b) Time-averaged CFI, $\overline{\mathcal{F}}_{C}$, over $n{=}10$ as a function of $L$ in DTC phase, $\omega{=}10^{-4}\pi/2$, and at the transition points, $\omega{=}\omega_{\max}$. Inset is  $\overline{\mathcal{F}}_C$, and time-averaged QFI, $\overline{\mathcal{F}}_Q$, as a function of $\omega$ for $L{=}20$.}\label{fig:Fig4}
\end{figure}

To provide a proof-of-principle demonstration of our DTC sensor, we propose superconducting quantum simulators, similar to Ref.~\cite{mi2022time}, for realizing our DTC. 
Rewriting the Floquet operator $U_{F}(\Omega,\varepsilon) = e^{-iH_P}e^{-i\Omega H_I}$ as 
\begin{equation}
    U_{F}(\Omega,\varepsilon) = \Big(\otimes_{j=1}^{L} e^{-i\Phi \sigma^x_j}\Big) \Big(\otimes_{j=1}^{L-1} e^{-ij\Omega\sigma^z_j\otimes\sigma^{z}_{j+1}}\Big), 
\end{equation}
clearly show that $U_F$ is an exact combinations of controlled-phase (CPHASE) gate $ZZ(j\Omega){=}e^{-ij\Omega \sigma^{z}_j\otimes \sigma^z_{j+1}}$  followed by single-qubit rotation $X(\Phi){=}e^{-i\Phi \sigma^{x}_j}$.
The proposed protocol, see Fig.~\ref{fig:Fig4} (a), for implementing our DTC includes three parts. After initializing the superconducting qubits in an arbitrary computational basis $|\mathbf{z}\rangle$, the Floquet operator $U_F$ is implemented identically $n$ times. 
Then at each cycle period the local polarization of each qubit $\langle \sigma^{z}_{j}\rangle$ is measured to track the overlap of the evolved state with $|\mathbf{z}\rangle$.
The newly developed tunable CPHASE gates allow us to engineer the gradient interaction between qubits to fulfill the requirement of our DTC. 
The technical details of the implementation and calibration of this two-qubit gate can be found in 
Ref.~\cite{mi2022time}.
In an array of $L$ qubits, implementing a single-period evolution $U_{F}$ requires $L{-}1$ two-qubit gates and $L$ single-qubit rotations.
The standard thermal relaxation time, $T_1$, and dephasing time, $T_2^*$, that characterize the typical operational time interval of the superconducting simulator have been reported as $T_1{\simeq}16$ $\mu$s and $T_2^*{\simeq}6$ $\mu$s~\cite{mi2022time}. 
Therefore, setting the CPHASE gate duration to ${\sim}20{+}20$ ns and the single-qubit gate duration to ${\sim}20$ ns allows us to excuse $n {=} T_2^*/0.06 {\simeq} 100$  cycling period in the operational interval of the circuit.
Our analysis shows that a simple configuration measurement described by projective operators as 
$\{\Pi_{\mathbf{z}}{=}|\mathbf{z}\rangle \langle \mathbf{z}|\}$ can saturate the quantum Cram\'{e}r-Rao bound. 
Here, $\{|\mathbf{z}\rangle\}$ represents the $2^L$ different computational basis. 
For $p_\mathbf{z}{=}|\langle \mathbf{z}|\psi_n \rangle|^2$ (with $|\psi_n \rangle{=} U_{F}^{n}(\Omega,\varepsilon)|\psi_0\rangle$) as the probability of finding the spins in the computational basis $|\mathbf{z}\rangle$, one can calculate the CFI. 
As it is clear from the inset of  Fig.~\ref{fig:Fig4} (b), the time-averaged CFI, $\overline{\mathcal{F}}_{C}{=}\frac{1}{N}\sum_{n=1}^{N}\mathcal{F}_{C}(nT)$, highly resembles the time-averaged QFI, $\overline{\mathcal{F}}_{Q}{=}\frac{1}{N}\sum_{n=1}^{N}\mathcal{F}_{Q}(nT)$. It shows two distinct behaviors in both DTC phase and non-DTC region and provides a sharp peak near the transition point $\omega_{\max}$. 
To extract the scaling behavior of the CFI, one can plot $\overline{\mathcal{F}}_{C}$ as a function of $L$ at different $\omega$.
Our results presented in Fig.~\ref{fig:Fig4} (b) show that the numerical simulations are properly mapped with a fitting function as $\overline{\mathcal{F}}_{C}{\propto}L^{\beta}$ with $\beta{=}3.158$ and $\beta{=}3.20$ for DTC phase and the transition point. 
This numerical simulation shows that the simple configuration measurement in our DTC sensor can saturate the quantum Cram\'{e}r-Rao bound and capture the quantum-enhanced sensitivity.

\section{Effect of Disordered Interaction}
\begin{figure}
    \centering
    \includegraphics[width=0.49\linewidth]{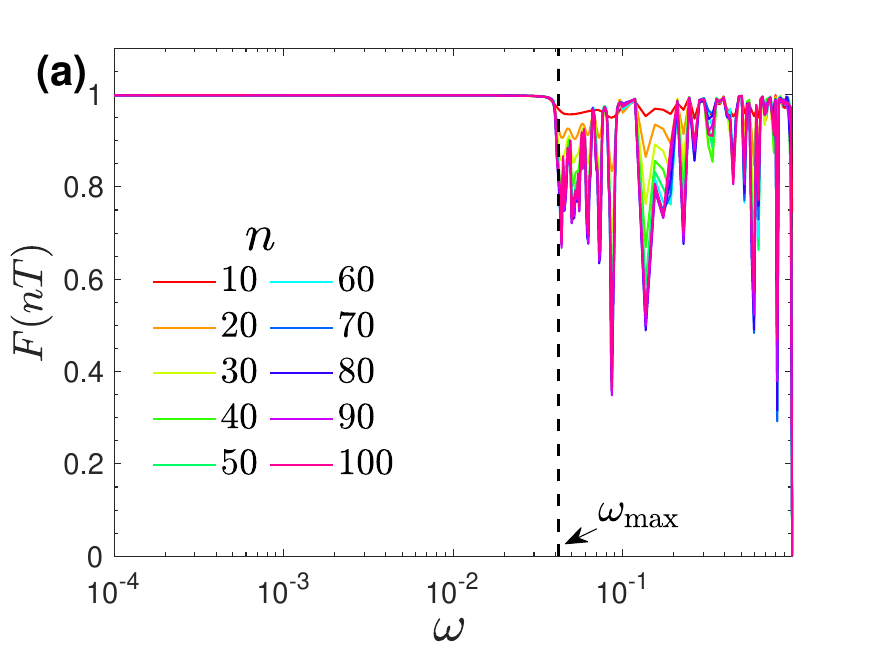}
    \includegraphics[width=0.49\linewidth]{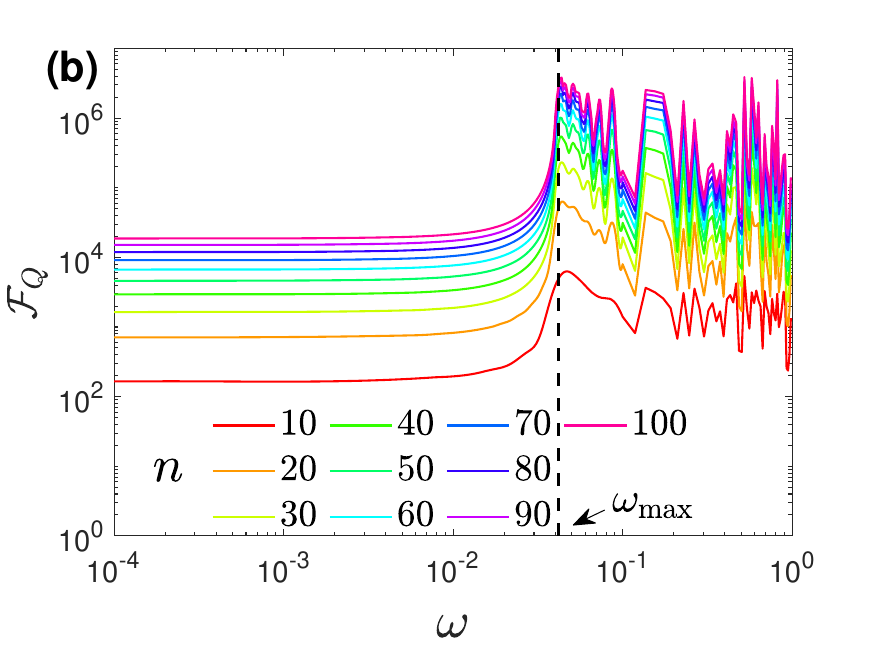}
    \includegraphics[width=0.49\linewidth]{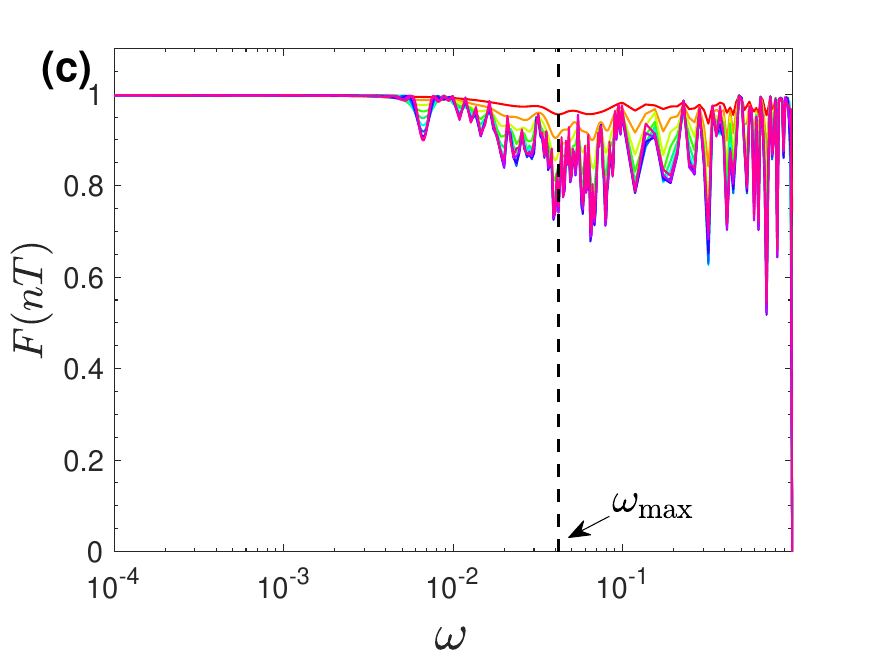}
    \includegraphics[width=0.49\linewidth]{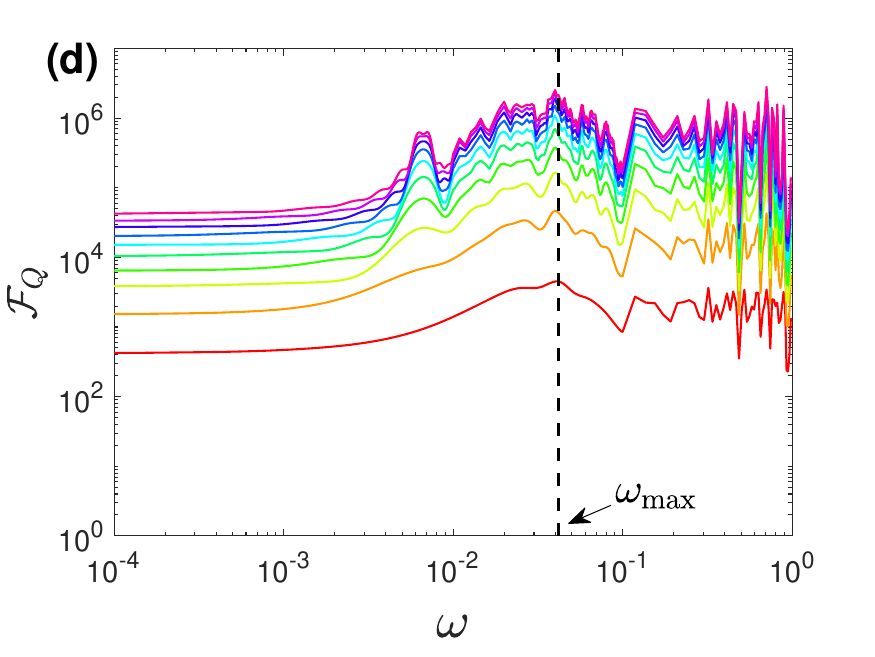}
    \caption{ (a) Dynamical behavior of the revival fidelity and (b) the QFI at stroboscopic times as a function of $\omega$ obtained for a disordered chain with strength $D{=}0.01$.
    (c) The revival fidelity and (d) the QFI obtained for a disordered chain while the disorder strength is $D{=}0.05$.  
    In all the panels the size of the 
    chain is $L{=}12$ and we set $\varepsilon{=}0.01$.}
    \label{fig:Fig6}
\end{figure}

So far, we have concentrated on a uniform gradient interaction among the system's components. In this section, we aim to analyze the effect of disorder in the gradient interaction. We replaced $H_I$ in Eq.~(\ref{Eq.Hamiltonian}) with $H_I{=}\sum_{j=1}^{L-1}j(1{-}d_j)\sigma^z_{j}\sigma^z_{j+1}$, where $d_j$ is drawn from a uniform distribution $[-D,D]$. 
The obtained results are presented in Fig.~\ref{fig:Fig6} for two values of the disorder strength namely, $D{=}0.01$ and $D{=}0.05$. As can be seen from Fig.~\ref{fig:Fig6} (a) and (b) a slight disorder in the gradient interaction can not affect the dynamical behavior of the DTC and, hence, its performance as a sensor. 
Increasing the disorder strength to $D{=}0.05$ slightly broadens the transition area, see Fig.~\ref{fig:Fig6} (c) and (d). Instead of a sharp transition from the DTC phase to the non-DTC region, some nontrivial oscillations blur the phase boundary. These results are obtained for a chain of size $L{=}12$ with $\varepsilon{=}0.01$ and have been averaged over $20$ repetition.   
This analysis shows that our DTC probe can tolerate a reasonable amount of disorder in the gradient interaction.

\section{Conclusion}
We have established a DTC with indefinite persistent oscillations and strong robustness against imperfections in the driving pulse. 
We show that this DTC can be used for measuring the coupling strength with quantum-enhanced precision over a region that extends over the entire DTC phase. Through finite-size scaling analysis, we characterize the nature of the phase transition as the second-order and determine relevant critical exponents.    
In addition, we show that imperfection in the pulse, enhances the precision before melting the DTC. The proposed mechanism is independent of the initial state and can be realized in existing quantum simulators.
It also shows robustness to slight disorder in the gradient interaction.

\section{Acknowledgement}
 We thank Fernando Iemini for insightful comments. A.B. acknowledges support from the National Natural
Science Foundation of China (Grants No. 12050410253,
No. 92065115, and No. 12274059), and the Ministry of Science and Technology of China (Grant No. QNJ2021167001L). R. Y. acknowledges support from the National Science Foundation of China for the International Young Scientists Fund (Grant No. 12250410242). This research was funded by the National Science Centre, Poland, Projects No. 2021/42/A/ST2/ 00017 (K.S.).

\appendix
\setcounter{section}{0}
\setcounter{equation}{0}
\setcounter{figure}{0}
\setcounter{table}{0}
\makeatletter
\renewcommand{\theequation}{A\arabic{equation}}

\section{Robustness of the  DTC against nonuniform imperfections }
In the main text, we analytically show that in the case of $\Omega{=}\pi/2$, our  DTC is robust against uniform imperfection defined as $\Phi=(1-\varepsilon)\frac{\pi}{2}$. 
In general, this imperfection can be nonuniform, namely it varies from site to site. 
In this scenario, the total imperfection $\Phi$ in $H_{P}$ (Eq.~(2) of the main text) can be replaced by $\Phi_{j}=(1-\varepsilon_{j})\frac{\pi}{2}$  with $\varepsilon_{j}$ as a random number which is selected from a uniform distribution as $[-\varepsilon,\varepsilon]$ with $\varepsilon{\neq}0$.  
To see the effect of this nonuniform imperfection on the  DTC in the case $\Omega{=}\pi/2$, we focus on the revival fidelity of a typical computational basis $|\mathbf{z}\rangle$.
Assume that $u_{\mathbf{z}}$ denotes the number of spins down in $|\mathbf{z}\rangle$, therefore the free evolution of the system governed by $H_{I}$ imposes a dynamical phase as $e^{-i\Omega H_{I}}|\mathbf{z}\rangle{=}(-i)^{L/2}(-1)^{u_{\mathbf{z}}} |\mathbf{z}\rangle$.
Then the first driving pulse evolves $|\mathbf{z}\rangle$ to a combination of $2^{L}$ computational basis, each with coefficient $\Pi_{j{\in}A}\Pi_{j^{'}\in\tilde{A}}(-i\cos(\varepsilon_{j})\sin(\varepsilon_{j^{'}}))$ wherein $A$  ($\tilde{A}$) are the collection of the unflipped (flipped) spins and $A{\cup}\tilde{A}{=}\{1,\cdots, L\}$.
Followed by the second period of evolution,  
one can show that $\langle \mathbf{z}|U_F^{2}(\Omega{=}\pi/2,\varepsilon{\neq}0)|\mathbf{z}\rangle$ is equal with the summation of $2^{L}$ choices of flipping $L$ spins with coefficient $(i)^{L}\Pi_{j{\in}A}\cos^{2}(\varepsilon_{j})\Pi_{j^{'}\in\tilde{A}}\sin^{2}(\varepsilon_{j^{'}})$. A straightforward simplification results in  $\langle \mathbf{z}|U_F^{2}(\Omega{=}\pi/2,\varepsilon{\neq}0)|\mathbf{z}\rangle{=}(- i)^{L}$  and, hence, $F(2T){=}1$. 
This calculation shows that regardless of the imperfections in the driving pulse, as long as $\Omega{=}\pi/2$ any initial state returns to itself after time $2T$, therefore, period-doubling oscillations of the revival fidelity resist indefinitely even in finite size systems. 
In the following, through an illustrative example, we provide more details on revival fidelity in a system of size  $L{=}4$ prepared initially in $|\mathbf{5}\rangle {=} |{\uparrow \downarrow \uparrow \downarrow}\rangle$. 
In fact we aim to calculate  $F(2T)=|\langle\mathbf{5}|U^2(\Omega{=}\pi/2,\varepsilon{\neq}0)|\mathbf{5}\rangle|^2$.
Note that here $c_j$ and $s_j$ are abbreviations for  $\cos(\Phi_{j})$ and $\sin(\Phi_{j})$, respectively.
\begin{align} 
e^{-i\Omega H_{I}}|\mathbf{5}\rangle =& (-i)^{L/2}(-1)^2 |\mathbf{5}\rangle
\end{align}
\begin{widetext}
\begin{align}    
e^{-iH_{p}}e^{-i\Omega H_{I}}|\mathbf{5}\rangle =&
(-i)^{L/2} \Big\{
c_1 c_2 c_3 c_4 |\mathbf{5}\rangle +
(-i) c_1 c_2 c_3 s_4 |\mathbf{4}\rangle +
(-i) c_1 c_2 s_3 c_4 |\mathbf{7}\rangle 
\cr
+&(-i)^2 c_1 c_2 s_3 s_4 |\mathbf{6}\rangle 
+ (-i)c_1 s_2 c_3 c_4 |\mathbf{1}\rangle +
(-i)^2 c_1 s_2 c_3 s_4 |\mathbf{0}\rangle
\cr
+& (-i)^2 c_1 s_2 s_3 c_4 |\mathbf{3}\rangle+
(-i)^3 c_1 s_2 s_3 s_4 |\mathbf{2}\rangle +  
(-i) s_1 c_2 c_3 c_4 |\mathbf{13}\rangle 
\cr
+& 
(-i)^2 s_1 c_2 c_3 s_4 |\mathbf{12}\rangle +
(-i)^2 s_1 c_2 s_3 c_4 |\mathbf{15}\rangle+
(-i)^3 s_1 c_2 s_3 s_4 |\mathbf{14}\rangle
\cr
+&
(-i)^2 s_1 s_2 c_3 c_4 |\mathbf{9}\rangle +
(-i)^3 s_1 s_2 c_3 s_4 |\mathbf{8}\rangle +
(-i)^3 s_1 s_2 s_3 c_4 |\mathbf{11}\rangle 
\cr
+&
(-i)^4 s_1 s_2 s_3 s_4 |\mathbf{10}\rangle 
\Big\}
\end{align}

\begin{align} 
e^{-i\Omega H_{I}}e^{-iH_{p}}e^{-i\Omega H_{I}}|\mathbf{5}\rangle =&
(-i)^{L} \Big\{
(-1)^2 c_1 c_2 c_3 c_4 |\mathbf{5}\rangle +
(-1)^1 (-i) c_1 c_2 c_3 s_4 |\mathbf{4}\rangle 
\cr +&
(-1)^3 (-i) c_1 c_2 s_3 c_4 |\mathbf{7}\rangle +
(-1)^2 (-i)^2 c_1 c_2 s_3 s_4 |\mathbf{6}\rangle 
\cr +&
(-1)^1 (-i) c_1 s_2 c_3 c_4 |\mathbf{1}\rangle +
(-1)^0 (-i)^2 c_1 s_2 c_3 s_4 |\mathbf{0}\rangle
\cr +&
(-1)^2 (-i)^2 c_1 s_2 s_3 c_4 |\mathbf{3}\rangle+
(-1)^1 (-i)^3 c_1 s_2 s_3 s_4 |\mathbf{2}\rangle  
\cr +&
(-1)^3 (-i) s_1 c_2 c_3 c_4 |\mathbf{13}\rangle +
(-1)^2 (-i)^2 s_1 c_2 c_3 s_4 |\mathbf{12}\rangle 
\cr +& 
(-1)^4 (-i)^2 s_1 c_2 s_3 c_4 |\mathbf{15}\rangle+
(-1)^3 (-i)^3 s_1 c_2 s_3 s_4 |\mathbf{14}\rangle
\cr +&
(-1)^2 (-i)^2 s_1 s_2 c_3 c_4 |\mathbf{9}\rangle +
(-1)^1 (-i)^3 s_1 s_2 c_3 s_4 |\mathbf{8}\rangle 
\cr +& 
(-1)^3 (-i)^3 s_1 s_2 s_3 c_4 |\mathbf{11}\rangle +
(-1)^2 (-i)^4 s_1 s_2 s_3 s_4 |\mathbf{10}\rangle 
\Big\}
\end{align}
\begin{align} 
e^{-iH_{p}}e^{-i\Omega H_{I}}e^{-iH_{p}}e^{-i\Omega H_{I}}|\mathbf{5}\rangle =&
(-i)^{L} \Big\{
(c_1 c_2 c_3 c_4)^2 +
(c_1 c_2 c_3 s_4)^2 +
(c_1 c_2 s_3 c_4)^2 +
(c_1 c_2 s_3 s_4)^2 + 
(c_1 s_2 c_3 c_4)^2 
\cr +&
(c_1 s_2 c_3 s_4)^2 +
(c_1 s_2 s_3 c_4)^2 +
(c_1 s_2 s_3 s_4)^2 +
(s_1 c_2 c_3 c_4)^2 
\cr +&
(s_1 c_2 c_3 s_4)^2 +
(s_1 c_2 s_3 c_4)^2 +
(s_1 c_2 s_3 s_4)^2 +
(s_1 s_2 c_3 c_4)^2 
\cr +&
(s_1 s_2 c_3 s_4)^2 +
(s_1 s_2 s_3 c_4)^2 +
(s_1 s_2 s_3 s_4)^2 
\Big\}
|\mathbf{5}\rangle+\cdots 
\end{align}
\begin{align} 
e^{-iH_{p}}e^{-i\Omega H_{I}}e^{-iH_{p}}e^{-i\Omega H_{I}}|\mathbf{5}\rangle =& 
(-i)^{L} |\mathbf{5}\rangle+\cdots 
\end{align}
\end{widetext}
Therefore, one has $F(2T)=|\langle\mathbf{5}|U^2(\Omega{=}\pi/2,\varepsilon{\neq}0)|\mathbf{5}\rangle|^2=1$.

\section{Imperfection effect and role of the initial state}
\begin{figure}
    \centering
    \includegraphics[width=0.49\linewidth]{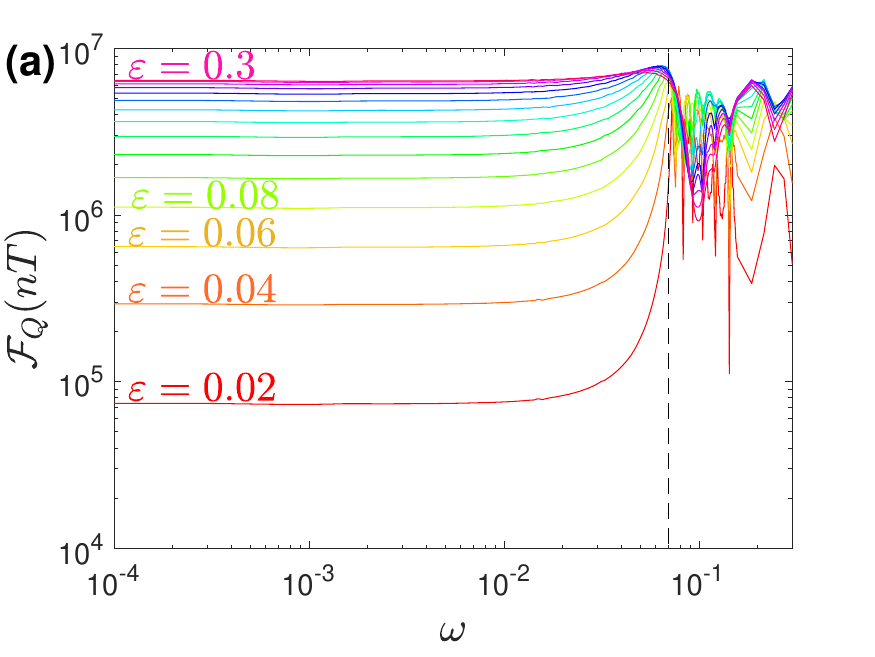}
    \includegraphics[width=0.49\linewidth]{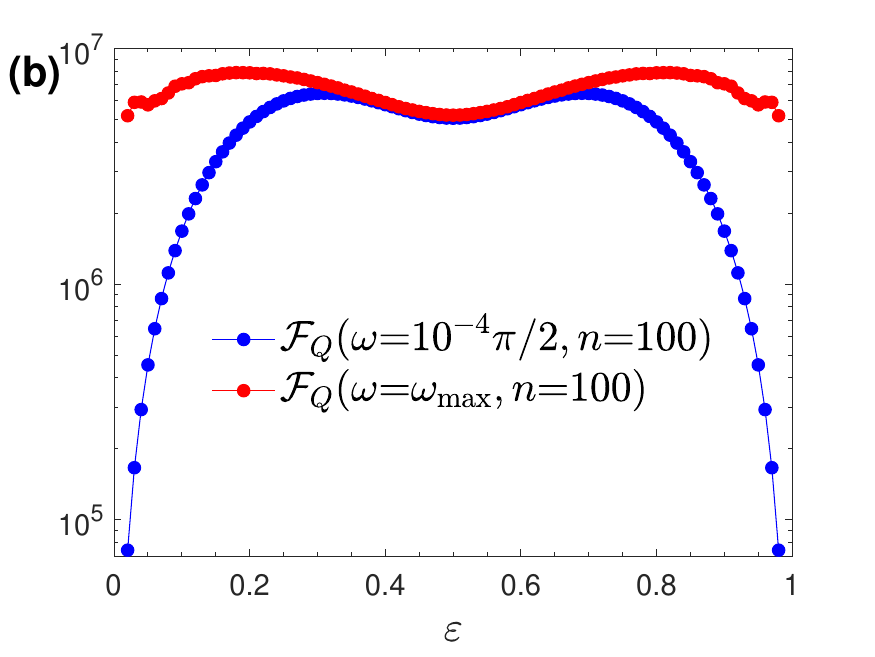}
    \caption{(a) The QFI versus $\omega$ for different $\varepsilon$. The dashed line determines the onset of the phase transition. (b) The QFI versus $\varepsilon$ for two values of $\omega$. The results are obtained after $n{=}100$ cycling periods in a system of size $L{=}12$, initialized in $|\psi_{0}\rangle{=}|\mathbf{0}\rangle{=}|{\uparrow,\uparrow,\cdots,\uparrow}\rangle$.}\label{fig:FigS1}
\end{figure}
\begin{figure}
    \centering
    \includegraphics[width=0.49\linewidth]{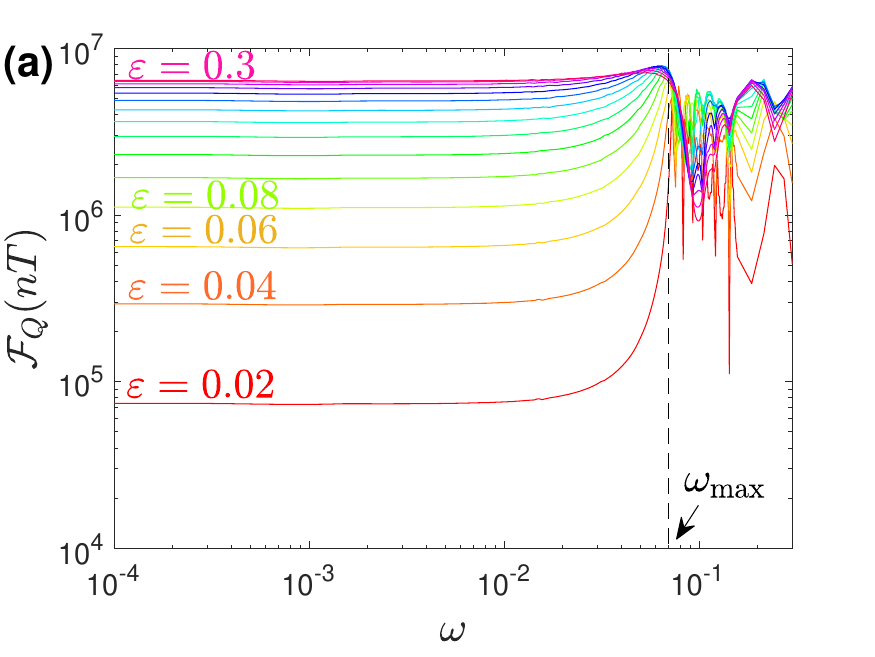}
    \includegraphics[width=0.49\linewidth]{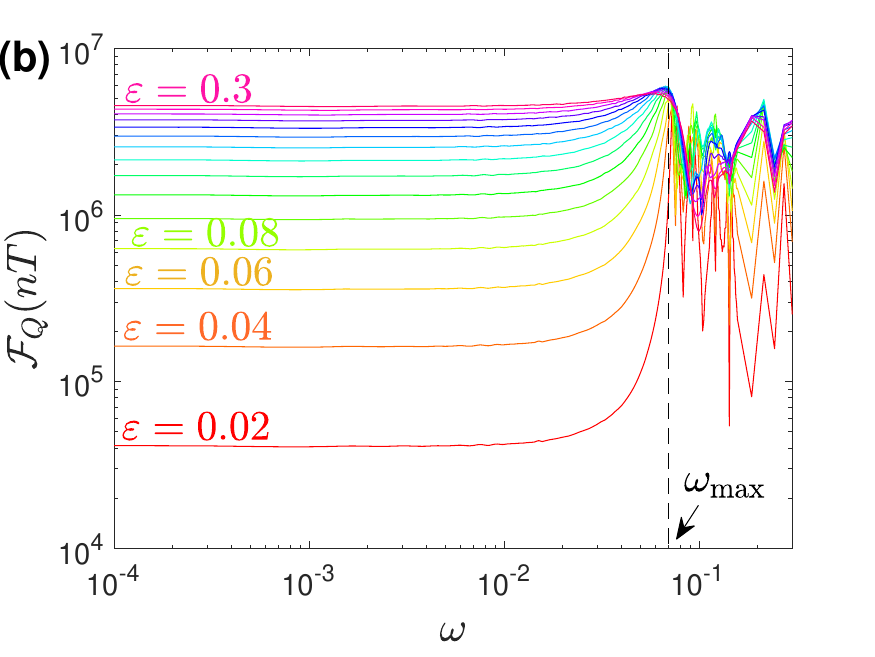}
    \caption{ (a)-(b) The QFI after $n{=}100$ period cycles as a function of deviation $\omega$ obtained for different values of $\varepsilon{\in}\{0.02,0.04,\cdots,0.3\}$ in a chain of size $L{=}12$ that is initialized in N\'{e}el state $|\psi_{0}\rangle{=}|\mathbf{1365}\rangle{=}|{\uparrow,\downarrow,\cdots,\uparrow,\downarrow}\rangle$, and a random state, respectively. }
    \label{fig:FigS2}
\end{figure}

In the main text, we analytically proved that our DTC is robust against uniform imperfection $\varepsilon$ in the driving pulse when $\omega{=}0$.
In the case of nonzero $\omega$, the situation becomes even more interesting. In Fig.~\ref{fig:FigS1}(a), we plot the QFI versus $\omega$ under driving pulse with various imperfections $\varepsilon$. 
While the qualitative behavior of the probe in the DTC phase is not affected by imperfection, increasing  $\varepsilon$ enhances the QFI. 
This can be understood as imperfect rotating pulses through involving a larger sector of the Hilbert space in the dynamics of the system imprints more information about $\omega$ into the quantum state. 
Notably, the onset of the transition from DTC to non-DTC region, which reflects a peak in $\mathcal{F}_{Q}$, is almost independent of the imperfection value.  
To assess the performance in a wider range of the imperfection, in Fig.~\ref{fig:FigS1}(b), $\mathcal{F}_{Q}$ as a function of $\varepsilon$ for deep inside the DTC phase, i.e. $\omega{=}10^{-4}\pi/2$, and at the transition point, i.e. $\omega{=}\omega_{\max}$, have been reported. 
The enhancement in the DTC phase, where the system is supposed to be strongly localized, has a remarkably stronger effect than at the transition point which already has features of both thermalization and localization. 
This interesting result is in sharp contrast with the usual sensors where the imperfections deteriorate the sensing power.
The results are obtained after $n{=}100$ cycling period for a chain of size $L{=}12$ initialized in $|\psi_{0}\rangle{=}|\mathbf{0}\rangle{=}|{\uparrow,\uparrow,\cdots,\uparrow}\rangle$.
\\
As we mention in the main text, the observed behavior of the DTC sensor concerning $\omega$ and $\varepsilon$ is general and independent of the initial states. 
To support this claim, in Figs.~\ref{fig:FigS2} (a)-(b), we depict the QFI $\mathcal{F}_{Q}(nT)$ as a function of $\omega$ in a system initialized in the N\'{e}el state $|\psi_{0}\rangle{=}|{\uparrow,\downarrow\cdots,\uparrow,\downarrow}\rangle$
and a random state, respectively.
The results are obtained 
after $n{=}100$ period cycles in a system of size $L{=}12$.
Curves with different colors correspond to different values of imperfections $\varepsilon\in\{0.02,\cdots,0.3\}$. 
In terms of $\omega$, one can see that the distinctive behavior of our system in both  DTC and non-DTC phases, as well as the transition between them at $\omega{=}\omega_{\max}$ reflects itself in all the considered initial states. 
Regarding the imperfection effect, by increasing $\varepsilon$, more information about $\omega$ can be printed in the evolved state, resulting in higher values of the QFI.
As is clear from Fig.~\ref{fig:FigS2}, this behavior is qualitatively independent of the initial states.

\section{ Melting transition of the DTC}
\begin{figure}
    \centering
    \includegraphics[width=0.49\linewidth]{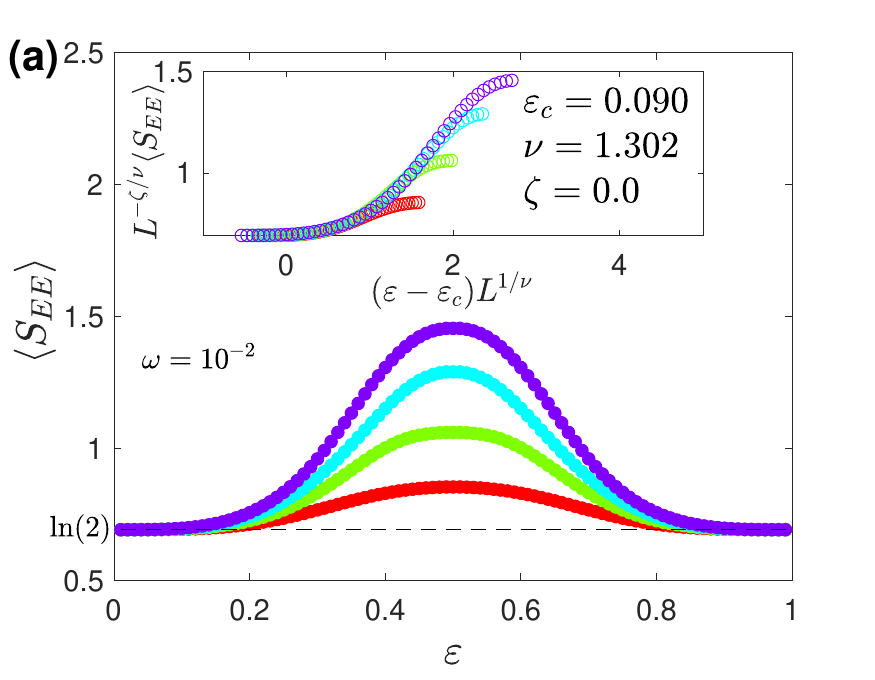}
    \includegraphics[width=0.49\linewidth]{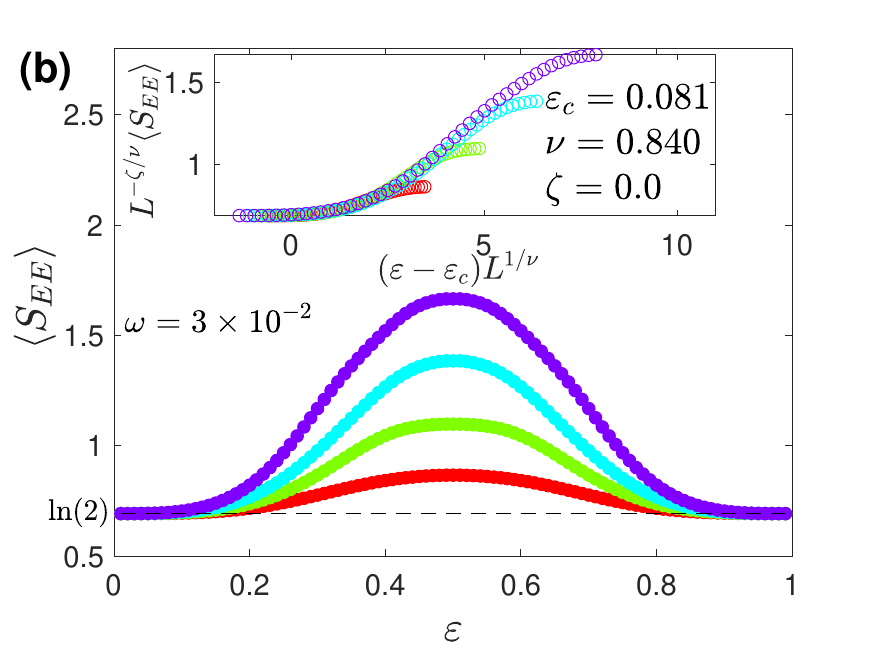}
    \includegraphics[width=0.49\linewidth]{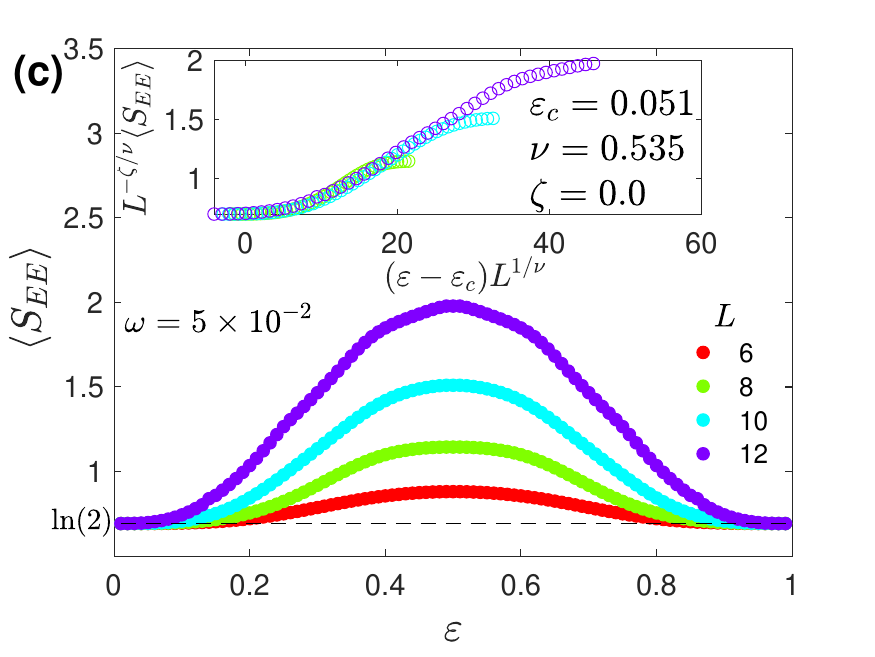}
    \includegraphics[width=0.49\linewidth]{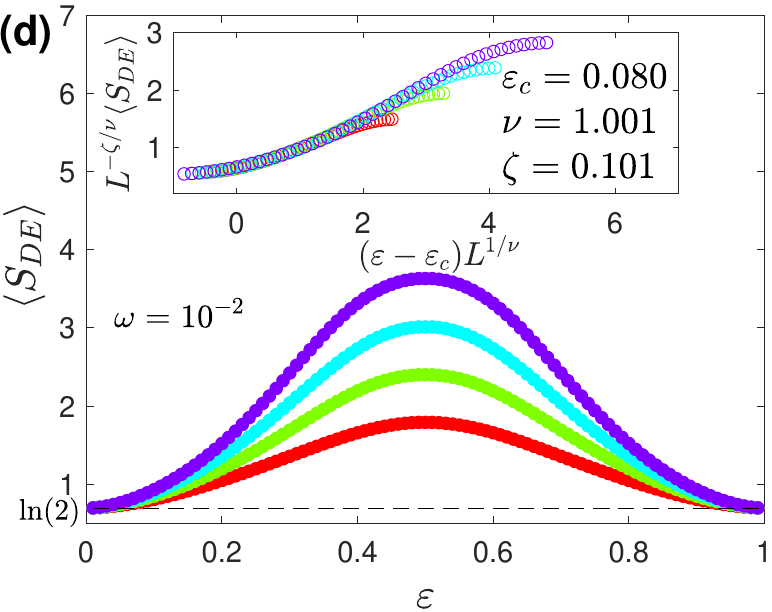}
    \includegraphics[width=0.49\linewidth]{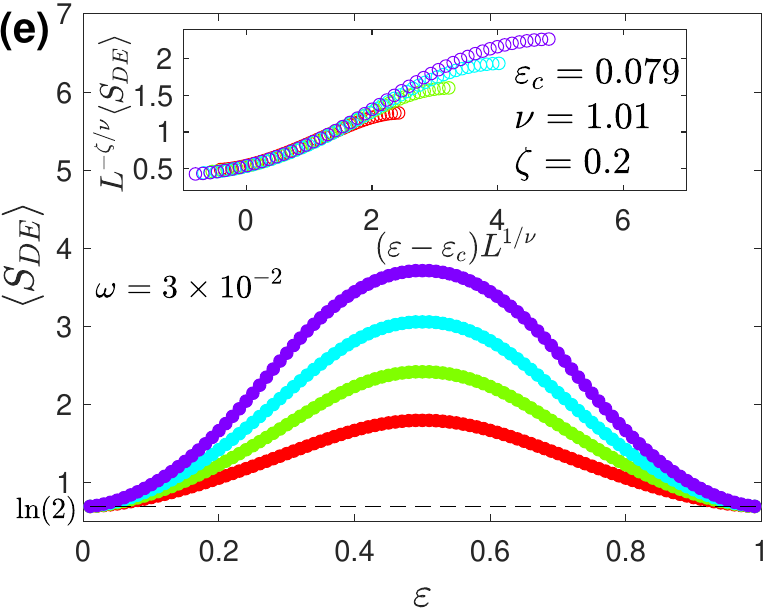}
    \includegraphics[width=0.49\linewidth]{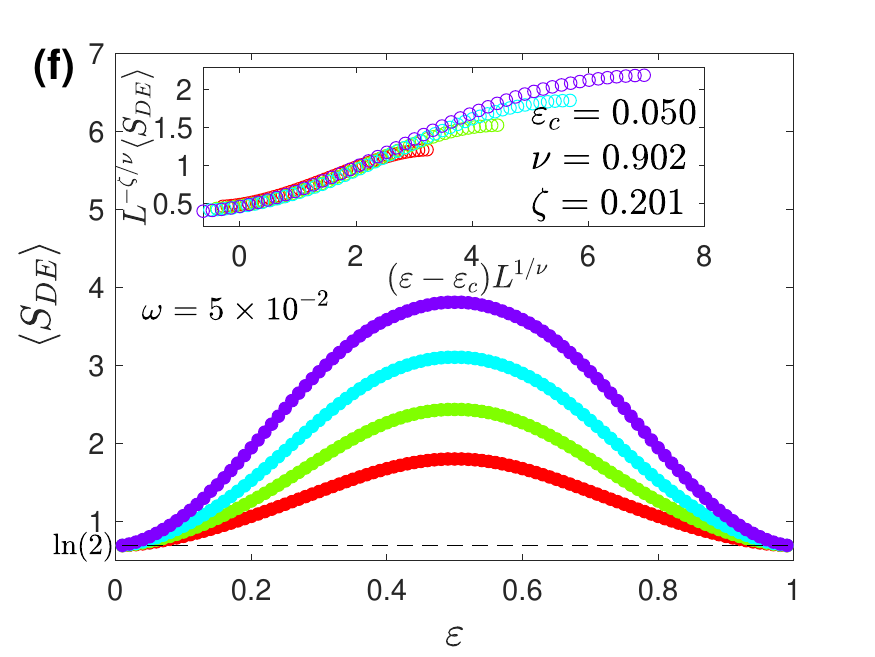}
    \caption{(a)-(c) the averaged entanglement entropy $\langle S_{EE} \rangle$ as a function of pulse imperfection $\varepsilon$ obtained for systems of different sizes and different $\omega$'s.
    (d)-(f) The averaged diagonal entropy $\langle S_{DE} \rangle$ as a function of  $\varepsilon$ in systems with various size and $\omega$'s. 
    The insets of the panels show the results of the finite-size scaling analysis and optimal data collapse of curves with different sizes, which happen for the reported critical properties $(\varepsilon_c, \nu, \zeta)$.}
    \label{fig:FigS3}
\end{figure}

In the main text, we show how increasing imperfection in the rotating pulse $\varepsilon$ can onset a phase transition between stable DTC and non-DTC phase. The transition driven by $\varepsilon$ is diagnosed by averaged entanglement entropy $\langle S_{EE} \rangle$ and averaged diagonal entropy $\langle S_{DE} \rangle$ obtained for all Floquet states of $U_{F}(\Omega,\varepsilon){=}\sum_{k=1}^{2^L}e^{-i\phi_k}|\phi_{k}\rangle\langle \phi_k|$. 
In the DTC phase, each Floquet state $|\phi_{k}\rangle$ is a maximally entangled GHZ state of a pair of computational basis states. 
For instance for small values of $\varepsilon$ one approximately has $|\phi_{1}\rangle{\cong}\frac{1}{\sqrt{2}}(|\mathbf{0}\rangle + |\mathbf{2^{L}-1}\rangle)$ and $|\phi_{2^L}\rangle{\cong}\frac{1}{\sqrt{2}}(|\mathbf{0}\rangle - |\mathbf{2^{L}-1}\rangle)$ with the corresponding eigenvalues as $\phi_{1}{\cong}E_{\mathbf{0}}$ and $\phi_{2^L}{\cong}E_{\mathbf{0}}\pm \pi$. 
Note that $\{E_{\mathbf{z}}\}$ are the eigenvalues of $H_{I}$ with $E_{\mathbf{z}}{=}E_{2^{L}-1-\mathbf{z}}$.
Clearly, deep inside the DTC regime, the entanglement entropy is $S_{EE}^{(k)}{\cong}\ln2$ for all $k$'s. 
This can be seen in Fig.~\ref{fig:FigS3} (a)-(c) which depict the averaged entanglement entropy $\langle S_{EE} \rangle$ for systems of various sizes and $\omega\in\{1,3,5\}\times10^{-2}$.
In this regime, by enlarging $\varepsilon$ the averaged entanglement entropy gets distance from $\ln2$ and peaks at its size- and $\omega$-dependent location, happening for $\varepsilon{=}0.5$. 
The behavior of the entanglement entropy concerning $L$ hints that the melting transition is of second-order type. This means that one can extract the critical properties for the transition by implementing finite-size scaling analysis.
However, as $\omega_{\max}{\propto}L^{-1}$, by increasing the system size the range of $\omega$'s that the DTC phase is stable for them, namely $\omega{<}\omega_{\max}$, shrinks. 
Therefore, the results for finite-size scaling analysis obtained using probes that for any given $\omega{<}\omega_{\max}$ these systems are within the DTC phase when $\varepsilon{\simeq}0$. 
Here, the numerical restriction in the ED method limits us to the system up to $L{=}12$.
Presuming that the averaged entanglement entropy follows an ansatz as $\langle S_{EE} \rangle=L^{\zeta/\nu}\mathcal{D}(L^{1/\nu}(\varepsilon-\varepsilon_{c}))$, then plotting $L^{-\zeta/\nu}\langle S_{EE} \rangle$ as a function of $L^{1/\nu}(\varepsilon-\varepsilon_{c})$ collapses the curves of different sizes. 
Here, $\zeta$ and $\nu$ are the critical exponents, $\varepsilon_{c}$ is the critical point, and $\mathcal{D}$ is an arbitrary function.
The best data collapse can be obtained for the optimal critical parameters $(\varepsilon_{c},\zeta,\nu)$.
In the insets of Fig.~\ref{fig:FigS3}(a)-(c), we present the best data collapse of the corresponding curves obtained for the reported critical parameters.
For the sake of completeness, we repeat the analysis above for the averaged diagonal entropy. 
The diagonal entropy contains partial information about the system by setting the off-diagonal terms of the half-system reduced density matrix to zero. 
In Fig.~\ref{fig:FigS3}(d)-(f), we present the obtained $\langle S_{DE} \rangle$ for various sizes and $\omega$'s.
Indeed, the behavior of the averaged diagonal entropy is qualitatively close to the averaged entanglement entropy in both  DTC and non-DTC phases. 
In particular, for small values of $\varepsilon$, one has $S_{EE}^{(k)}=S_{DE}^{(k)}=\ln2$.
In the insets of Fig.~\ref{fig:FigS3}(d)-(f), we present the results of the finite-size scaling analysis that we established for this quantity.
Surprisingly the obtained $\varepsilon_{c}$ for both $\langle S_{EE} \rangle$ and $\langle S_{DE} \rangle$ are close.
Note that in the transition between the DTC and non-DTC phases driven by $\varepsilon$, one may not observe quantum-enhanced sensitivity in the process of sensing $\varepsilon$.

%

\end{document}